\documentclass[journal,twocolumn]{IEEEtran}
\usepackage[numbers,sort&compress]{natbib}
\usepackage{algorithm}
\usepackage{algpseudocode}
\usepackage{amsthm}
\usepackage{amsmath}
\usepackage{subfigure}
\usepackage{graphicx}
\usepackage{amsfonts}
\usepackage{stfloats}

\newtheorem{lemma}{Lemma}
\newtheorem{corollary}{Corollary}
\newtheorem{theorem}{Theorem}

\IEEEoverridecommandlockouts
\begin{document}

\title{FFT Algorithm for Binary Extension Finite Fields and its Application to Reed-Solomon Codes}

\author{Sian-Jheng Lin,~\IEEEmembership{Member,~IEEE},
Tareq Y. Al-Naffouri,~\IEEEmembership{Member,~IEEE}, and
Yunghsiang S. Han,~\IEEEmembership{Fellow,~IEEE}
\thanks{
This work was supported in part by CAS Pioneer Hundred Talents Program and the National Science of Council (NSC) of Taiwan under Grants NSC 102-2221-E-011-006-MY3, NSC 101-2221-E- 011-069-MY3. S.-J. Lin is with the School of Information Science and Technology, University of Science and Technology of China~ (USTC), Hefei, China and the Electrical Engineering Department, King Abdullah University of Science and Technology~(KAUST), Kingdom of Saudi Arabia~(e-mail: sjlin@ustc.edu.cn), Tareq Y. Al-Naffouri is with the Electrial Engineering Department at King Abdullah University of Science and Technology (KAUST), Thuwal, Makkah Province, Kingdom of Saudi Arabia. (e-mail: tareq.alnaffouri@kaust.edu.sa), and Y. Han is with the Department of Electrical Engineering, National Taiwan University of Science and Technology, Taipei, Taiwan. (e-mail: yshan@mail.ntust.edu.tw).
}}
\maketitle
\begin{abstract} 
Recently, a new polynomial basis over binary extension fields was proposed such that the fast Fourier transform (FFT) over such fields can be computed in the complexity of order $\mathcal{O}(n\lg(n))$, where $n$ is the number of points evaluated in FFT. In this work, we reformulate this FFT algorithm such that it can be easier understood and be extended to develop frequency-domain decoding algorithms for $(n=2^m,k)$ systematic Reed-Solomon~(RS) codes over $\mathbb{F}_{2^m},m\in \mathbb{Z}^+$, with $n-k$ a power of two.  First, the basis of syndrome polynomials is reformulated in the decoding procedure so that the new transforms can be applied to the decoding procedure. A fast extended Euclidean algorithm is developed to determine the error locator polynomial. The computational complexity of the proposed decoding algorithm is $\mathcal{O}(n\lg(n-k)+(n-k)\lg^2(n-k))$, improving upon the best currently available decoding complexity $\mathcal{O}(n\lg^2(n)\lg\lg(n))$, and reaching the best known complexity bound that was established by Justesen in 1976. However, Justesen's approach is only for the codes over some specific fields, which can apply Cooley-Tucky FFTs. As revealed by the computer simulations, the proposed decoding algorithm is $50$ times faster than the conventional one for the $(2^{16},2^{15})$ RS code over $\mathbb{F}_{2^{16}}$.
\end{abstract}
\thispagestyle{plain}
\pagestyle{plain}
\setcounter{page}{1}

\section{Introduction}
Reed-Solomon (RS) codes are a class of block error-correcting codes that were invented by Reed and Solomon~\cite{1960} in 1960. An $(n,k)$ RS code is constructed over $\mathbb{F}_{q}$, for $n=q-1$. Its extended version, called extended Reed-Solomon codes~\cite{6769771}, admits a codeword length of up to $n=q$ or $n=q+1$. The systematic version of $(n,k)$ RS code appends $n-k$ parity symbols to the $k$ message symbols, forming a codeword of length $n$. RS codes are maximum distance separable~(MDS). $(n,k)$ RS codes can correct up to $\lfloor (n-k)/2\rfloor$ erroneous symbols. Nowadays, RS codes have numerous important applications, including  barcodes~(such as QR codes), storage devices~(such as Blu-ray Discs), digital television~(such as DVB and ATSC), and data transmission technologies~(such as DSL and WiMAX). RS codes are also used to design other forward error correction codes, such as regenerating codes~\cite{5961826}\cite{6977928} and local reconstruction codes~\cite{Huang:2012}. The wide range of applications of RS codes raises an important issue concerning their computational complexity. More specifically, since the practical implementations of RS codes are typically over binary extension finite fields, the complexity of RS codes over those fields has received more attentions than that over others~\cite{Chen:2008}\cite{1599587}.

The conventional syndrome-based RS decoding algorithm has quadratic complexities. Some fast approaches~\cite{1055516}\cite{Gao02anew} are based on FFTs or fast polynomial arithmetic techniques. However, the structures of FFTs over finite fields vary with the sizes of fields $\mathbb{F}_q$. When $q-1$ is a smooth number, meaning that $q-1$ can be factorized into many small primes, the Cooley-Tucky FFT in $\mathcal{O}(n\lg(n))$ field additions and field multiplications can be applied. A conventional case involves choosing Fermat primes $q\in\{2^m+1|m=1,2,4,8,16\}$. Based on such FFTs, Justesen~\cite{1055516} gave an $\mathcal{O}(n\lg^2(n))$ approach for decoding $(n,k)$ RS code over $\mathbb{F}_{2^m+1}$. Another approach to solve the key equations of BCH codes was proposed by Pan~\cite{Pan:1997}, and it reduces a factor of $\lg n$ when the characteristic of the field is large enough. However, the algorithm~\cite{Pan:1997} does not have improvement for the codes over binary extension fields. If $q-1$ is not smooth, Cooley-Tucky FFTs are inapplicable. In this case, the FFTs over arbitrary fields~\cite{Schonhage1977}\cite{Cantor1991} can be applied and it requires $\mathcal{O}(n\lg(n)\lg\lg (n))$ field operations. Gao~\cite{Gao02anew} presented an $\mathcal{O}(n\lg^2(n)\lg\lg(n))$ RS decoding algorithm over arbitrary fields, by utilizing fast polynomial multiplications~\cite{Gathen:2013}. Further, for the codes over $\mathbb{F}_{2^m}$, the additive FFT~\cite{5625613}, that requires $\mathcal{O}(n\lg(n)\lg\lg(n))$ operations, can be applied to reduce the leading constant further. To authors' knowledge, the additive FFT~\cite{5625613} is the fastest algorithm over $\mathbb{F}_{2^m}$ so far.

As RS codes are typically constructed over binary extension fields, we consider this case in this paper. Clearly, if one wants to remove the extra factor $\lg\lg(n)$ in the RS algorithms over binary extension fields, the FFTs in $\mathcal{O}(n\lg (n))$ are required. Recently, Lin et al.~\cite{4319038} showed a new way to solve aforementioned FFT problem. The paper~\cite{4319038} defined a new polynomial basis based on subspace polynomials over $\mathbb{F}_{2^m}$. For a polynomial of degree less than $h$ in this new basis, the $h$-point multipoint evaluations can be made in $\mathcal{O}(h\lg(h))$ field operations. Based on the multipoint evaluation algorithm, encoding/erasure decoding algorithms for $(n,k)$ RS codes~\cite{4319038} were proposed to achieve $\mathcal{O}(n\lg(n))$. However, the error-correction RS decoding algorithm based on the new basis was not yet provided.

This paper develops an error correction decoding algorithm for $(n=2^m,k)$ RS codes over $\mathbb{F}_{2^m}$, for $k/n\geq 0.5$ and $(n-k)$ a power of two.\footnote{There are many $(n,k)$ can be chosen when $n=2^m,k=2^m-2^t$, where $t<m$. } In practice RS codes usually have rates $k/n \geq 0.5$. The complexity of the proposed algorithm is given by $\mathcal{O}(n\lg(n-k)+(n-k)\lg^2(n-k))$. Holding constant the code rate $k/n$ yields a complexity $\mathcal{O}(n\lg^2(n))$, which is better than the best existing complexity of $\mathcal{O}(n\lg^2(n)\lg\lg(n))$, that was achieved by Gao~\cite{Gao02anew} in 2002. The algorithm is based on the non-standard polynomial basis~\cite{4319038}. To embed the new basis into the decoding algorithm, we reformulate the decoding formulas such that all arithmetics are performed on the new basis. The key equation is solved by the Euclidean algorithm, and thus the fast polynomial divisions, as well as the Euclidean algorithm in the new basis are proposed. Finally, we combine those algorithms, resulting in a fast error-correction RS decoding algorithm. The major contributions of this paper are summarized as follows.
\begin{enumerate}
\item An alternative description of the algorithms~\cite{4319038} for the new polynomial basis is presented.
\item An $\mathcal{O}(h\lg(h))$ fast polynomial division in the new basis is derived.
\item An $\mathcal{O}(h\lg^2(h))$ fast half-GCD algorithm in the new basis is presented.
\item An $\mathcal{O}(n\lg(n-k))$ RS encoding algorithm is presented, for $n-k$ a power of two.
\item A syndrome-based RS decoding algorithm that is based on the new basis is demonstrated.
\item An $\mathcal{O}(n\lg(n-k)+(n-k)\lg^2(n-k))$ RS decoding algorithm is presented, for $n-k$ a power of two.
\end{enumerate}
Notably, \cite{4319038} gave the encoding algorithms for RS codes with the complexity $O(n\lg (k))$, for $k$ a power of two. The encoding algorithm~\cite{4319038} is suitable for coding rate $k/n\leq 0.5$; however, the proposed encoding algorithm in this work is suitable for $k/n\geq 0.5$.

The rest of this paper is organized as follows. Section \ref{sec:fastTransform} reviews the definitions of the polynomial basis. The multipoint evaluation algorithm is provided in Sec. \ref{sec:polyEva2}. Section \ref{sec:polynomialDivision} provides an alternative polynomial basis that is constructed using monic polynomials. The polynomial operations that are used in the encoding/decoding of RS codes are explicated. Section~\ref{sec:Euclidean} presents the fast extended Euclidean algorithm that is based on the half-GCD method. Section \ref{sec:Encoding} and Section \ref{sec:Decoding} introduce the algorithms for encoding and decoding RS codes. Section \ref{sec:conclusion} presents simulations and draws conclusions. 

\section{Polynomial basis in $\mathbb{F}_{2^m}[x]/x^{2^m}-x$}\label{sec:fastTransform}
This section reviews the subspace polynomials over $\mathbb{F}_{2^m}$, and the polynomial basis defined in~\cite{4319038}.

\subsection{Subspace polynomial}\label{sec:Arithmetic}
Let $\mathbb{F}_{2^m}$ denote an extension finite field with dimension $m$ over $\mathbb{F}_2$. Let $\overline{v}=(v_0,v_1,\dots ,v_{m-1})$ denote a basis of $\mathbb{F}_{2^m}$. That is, all $v_i\in \mathbb{F}_{2^m}$ are linearly independent over $\mathbb{F}_2$. A $k$-dimensional space $V_k$ of $\mathbb{F}_{2^m}$ is defined as
\begin{equation}\label{eq:V_k}
\begin{aligned}
&V_k=\mathrm{Span}(\overline{v}_k)\\
=&\{i_0\cdot v_0+i_1\cdot v_1+\dots +i_{k-1}\cdot v_{k-1}|\forall i_j\in \{0,1\} \},
\end{aligned}
\end{equation}
where $\overline{v}_k=(v_0,v_1,\dots ,v_{k-1})$ is a basis of space $V_k$, and $k\leq m$. We can form a strictly ascending chain of subspaces given by
\[
\{0\}=V_0\subset V_1\subset V_2\subset \dots \subset V_m=\mathbb{F}_{2^m}.
\]
Let $\{\omega_i\}_{i=0}^{2^m-1}$ denote the elements of $\mathbb{F}_{2^m}$. Each element is defined as
\[
\omega_i= i_0\cdot v_0+i_1\cdot v_1+\dots +i_{m-1}\cdot v_{m-1},
\]
where $i_j\in \{0,1\}$ is the binary representation of $i$. That is,
\[
i=i_0+i_1\cdot 2+\dots +i_{m-1}\cdot 2^{m-1}, \forall i_j\in \{0,1\}.
\]
This implies that $V_k=\{\omega_i\}_{i=0}^{2^k-1}$, for $k=0,1,\dots ,m$. Note that $\omega_0=0$ is the additive identity in the filed. In this work, $\omega_0$ and $0$ will be used interchangeably when there is no confusion.

The subspace polynomial \cite{ORE33,Cantor1989285,5625613} of $V_k$ is defined as
\begin{equation}\label{eq:w_j(x)}
s_k(x)=\prod_{a\in V_k}(x-a),
\end{equation}
and it is clear to see that $\deg(s_k(x))=2^k$. For example, $s_0(x)=x$, and $s_2(x)=x(x-v_0)(x-v_1)(x-v_0-v_1)$. The properties of $s_k(x)$ are given in \cite{ORE33,236882}.
\begin{theorem}[\cite{ORE33,236882}]\label{canonical_form_W_i(x)}
(i). $s_k(x)$ is an $\mathbb{F}_2$-linearlized polynomial for which
\begin{equation}\label{eq:W_i(x)}
s_k(x)=\sum_{i=0}^{k}s_{k,i}x^{2^i},
\end{equation}
with each $s_{k,i}\in \mathbb{F}_{2^m}$. This implies that
\begin{equation}\label{linear}
s_k(x+y)=s_k(x)+s_k(y), \forall x,y \in \mathbb{F}_{2^m}.
\end{equation}

(ii). The formal derivative of $s_k(x)$ is a constant
\begin{equation}
s'_k(x)=\prod_{a\in V_k\setminus \{0\}}a.
\end{equation}
\end{theorem}

The recursive form~\cite{ORE33} of subspace polynomials is given by
\begin{equation}\label{eq:w_j2(x)}
s_0(x)= x;
\end{equation}
\begin{equation}\label{eq:w_j2(x)-1}
\begin{aligned}
&s_j(x)= s_{j-1}(x)s_{j-1}(x-v_{j-1})\\
=&(s_{j-1}(x))^2-s_{j-1}(v_{j-1})s_{j-1}(x)\quad j=1,2,\dots , m.
\end{aligned}
\end{equation}

\subsection{Polynomial basis}\label{sec:NewPolynomialBasis}
Let $\mathbb{\bar{X}}=\{\bar{X}_0(x),\bar{X}_1(x),\dots ,\bar{X}_{2^m-1}(x)\}$ denote a basis of $\mathbb{F}_{2^m}[x]/(x^{2^m}-x)$. Each $\bar{X}_i(x)$ is defined as
\begin{equation}\label{eq:X_i(x)}
\bar{X}_i(x)={X_i(x)}/ {p_i},
\end{equation}
where
\begin{equation}\label{eq:barX_i(x)}
X_i(x)=\prod_{j=0}^{m-1}(s_j(x))^{i_j},\qquad p_i=\prod_{j=0}^{m-1}(s_j(v_j))^{i_j},
\end{equation}
and each $i_j\in \{0,1\}$ is the binary representation of $i$. Notice that $(s_j(x))^0=(s_j(v_j))^0=1$. For example, $\bar{X}_0(x)=1$, and $\bar{X}_3(x)={X_3(x)}/{p_3}=(s_0(x)s_1(x))/(s_0(v_0)s_1(v_1))$. It can be seen that $\deg(\bar{X}_i(x))=i$, and thus the basis $\mathbb{\bar{X}}$ can represent all elements in $\mathbb{F}_{2^m}[x]/(x^{2^m}-x)$.

A polynomial $\bar D_h(x)$ of degree $h$ in the basis $\mathbb{\bar{X}}$ is represented as
\begin{equation}\label{eq:D_h(x)}
\bar D_h(x)=\sum_{i=0}^{h-1}\bar d_i\bar{X}_i(x),
\end{equation}
with each $\bar d_i \in \mathbb{F}_{2^m}$. Throughout this paper, $\bar D_h=(\bar d_0, \bar d_1,\dots , \bar d_{h-1})
$ is used to indicate the vector of the coefficients of $\bar D_h(x)$. Due to the fact $\deg (\bar{X}_i(x))=i$, the new basis possesses the following properties.

\begin{corollary}\label{lemmacon0}
Given a polynomial $f(x)\in \mathbb{F}_{2^m}[x]/(x^{2^m}-x)$ respectively expressed in the monmial basis and $\mathbb{\bar{X}}$
\[
f(x)=\sum_{i=0}^{2^m-1} f_i^{(0)}x^i=\sum_{i=0}^{2^m-1} f_i^{(1)}\bar{X}_i(x),
\]
the following properties hold.
\begin{enumerate}
\item $f_i^{(0)}=f_i^{(1)}=0$, for $i\geq h+1$.
\item $f_h^{(1)}=f_h^{(0)}\cdot p_h$.
\item For $0\leq j \leq h$, $(f_j^{(1)}, f_{j+1}^{(1)},\dots ,f_h^{(1)})$ is determined by $(f_j^{(0)}, f_{j+1}^{(0)},\dots ,f_h^{(0)})$, and vice versa.
\end{enumerate}
\end{corollary}

\section{Multipoint evaluations at $V_k$}\label{sec:polyEva2}
\begin{algorithm}[t]
\caption{\label{alg:FFT} Transform of the basis $\bar{X}$}
\begin{algorithmic}[1]
\Require
$\mathrm{FFT}_\mathbb{\bar{X}}(\bar{D}_{2^k}, k, \beta)$: $\bar{D}_{2^k}=(\bar{d}_0,\bar{d}_1,\dots ,\bar{d}_{2^k-1})$, $k$ is the binary logarithm of size, and $\beta\in \mathbb{F}_{2^m}$
\Ensure
$2^k$ evaluations $\underline{D}_{2^k}=(\underline{d}_0,\underline{d}_1,\dots ,\underline{d}_{2^k-1})$, where each $\underline{d}_i=\bar{D}_{2^k}(\omega_i+\beta)$
\If {$k=0$} \Return $\bar{d}_0$
\EndIf

\For {$i=0,\dots ,2^{k-1}-1$}
\State $g_i^{(0)} \leftarrow \bar{d}_i+\frac{s_{k-1}(\beta)}{s_{k-1}(v_{k-1})}\bar{d}_{i+2^{k-1}}$
\State $g_i^{(1)} \leftarrow g_i^{(0)}+\bar{d}_{i+2^{k-1}}$
\EndFor
\State Call $V_0\leftarrow\mathrm{FFT}_\mathbb{\bar{X}}(\bar{D}_{2^{k-1}}^{(0)}, k-1, \beta)$, where $\bar{D}_{2^{k-1}}^{(0)}=(g_0^{(0)},\dots ,g_{2^{k-1}-1}^{(0)})$ and $V_0=(\underline{d}_0,\dots ,\underline{d}_{2^{k-1}-1})$
\State Call $V_1\leftarrow\mathrm{FFT}_\mathbb{\bar{X}}(\bar{D}_{2^{k-1}}^{(1)}, k-1, v_{k-1}+\beta)$, where $\bar{D}_{2^{k-1}}^{(1)}=(g_0^{(1)},\dots ,g_{2^{k-1}-1}^{(1)})$ and $V_1=(\underline{d}_{2^k},\dots ,\underline{d}_{2^k-1})$
\State \Return $\underline{D}_{2^k}=(\underline{d}_0,\underline{d}_1,\dots ,\underline{d}_{2^k-1})$
\end{algorithmic}
\end{algorithm}

\begin{algorithm}[t]
\caption{\label{alg:IFFT} Inverse transform of the basis $\bar{X}$}
\begin{algorithmic}[1]
\Require
$\mathrm{IFFT}_\mathbb{\bar{X}}(\underline{D}_{2^k}, k, \beta)$: $\underline{D}_{2^k}=(\underline{d}_0,\underline{d}_1,\dots ,\underline{d}_{2^k-1})$, where each $\underline{d}_i=\bar{D}_{2^k}(\omega_i+\beta)$, $k$ is the binary logarithm of size, and $\beta\in \mathbb{F}_{2^m}$
\Ensure
$\bar{D}_{2^k}=(\bar{d}_0,\bar{d}_1,\dots ,\bar{d}_{2^k-1})$, the coefficients of $\bar{D}_{2^k}(x)$

\If {$k=0$} \Return $\underline{d}_0$
\EndIf

\State Call $\bar{D}_{2^{k-1}}^{(0)}\leftarrow\mathrm{IFFT}_\mathbb{\bar{X}}(V_0, k-1, \beta)$, where $V_0=(\underline{d}_0,\dots ,\underline{d}_{2^{k-1}-1})$ and $\bar{D}_{2^{k-1}}^{(0)}=(g_0^{(0)},\dots ,g_{2^{k-1}-1}^{(0)})$

\State Call $\bar{D}_{2^{k-1}}^{(1)}\leftarrow\mathrm{IFFT}_\mathbb{\bar{X}}(V_1, k-1, v_{k-1}+\beta)$, where $V_1=(\underline{d}_{2^k},\dots ,\underline{d}_{2^k-1})$ and $\bar{D}_{2^{k-1}}^{(1)}=(g_0^{(1)},\dots ,g_{2^{k-1}-1}^{(1)})$

\For {$i=0,\dots ,2^{k-1}-1$}
\State $\bar{d}_{i+2^{k-1}} \leftarrow g_i^{(0)}+g_i^{(1)}$
\State $\bar{d}_i\leftarrow g_i^{(0)}+\frac{s_{k-1}(\beta)}{s_{k-1}(v_{k-1})}\bar{d}_{i+2^{k-1}}$
\EndFor

\State \Return $\bar{D}_{2^k}=(\bar{d}_0,\bar{d}_1,\dots ,\bar{d}_{2^k-1})$
\end{algorithmic}
\end{algorithm}

\begin{figure}
\center
\subfigure[The transform]{
   \includegraphics[width=1.0\columnwidth]{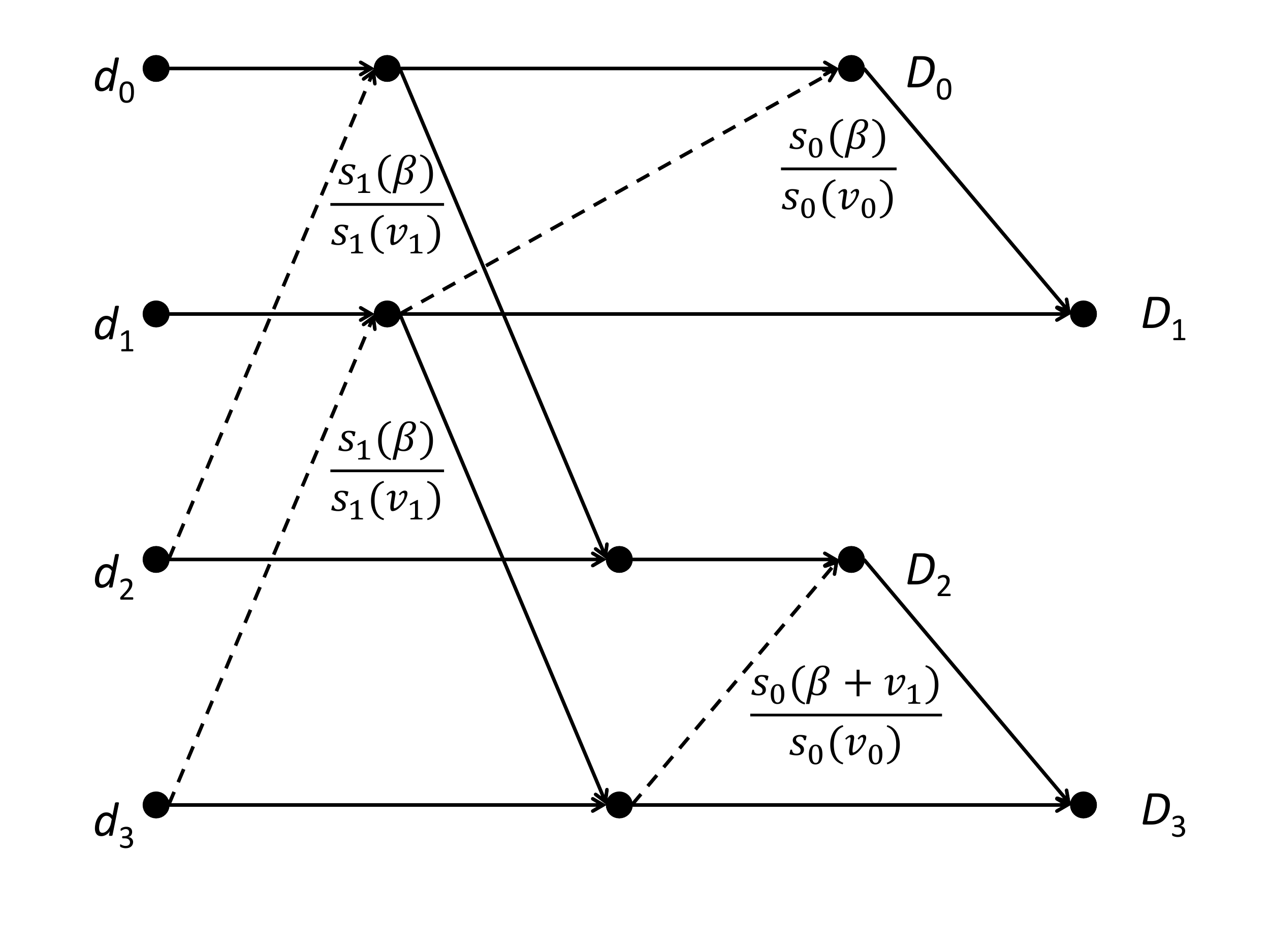}
   \label{fig:subfig1}
}

\subfigure[The inverse transform]{
   \includegraphics[width=1.0\columnwidth]{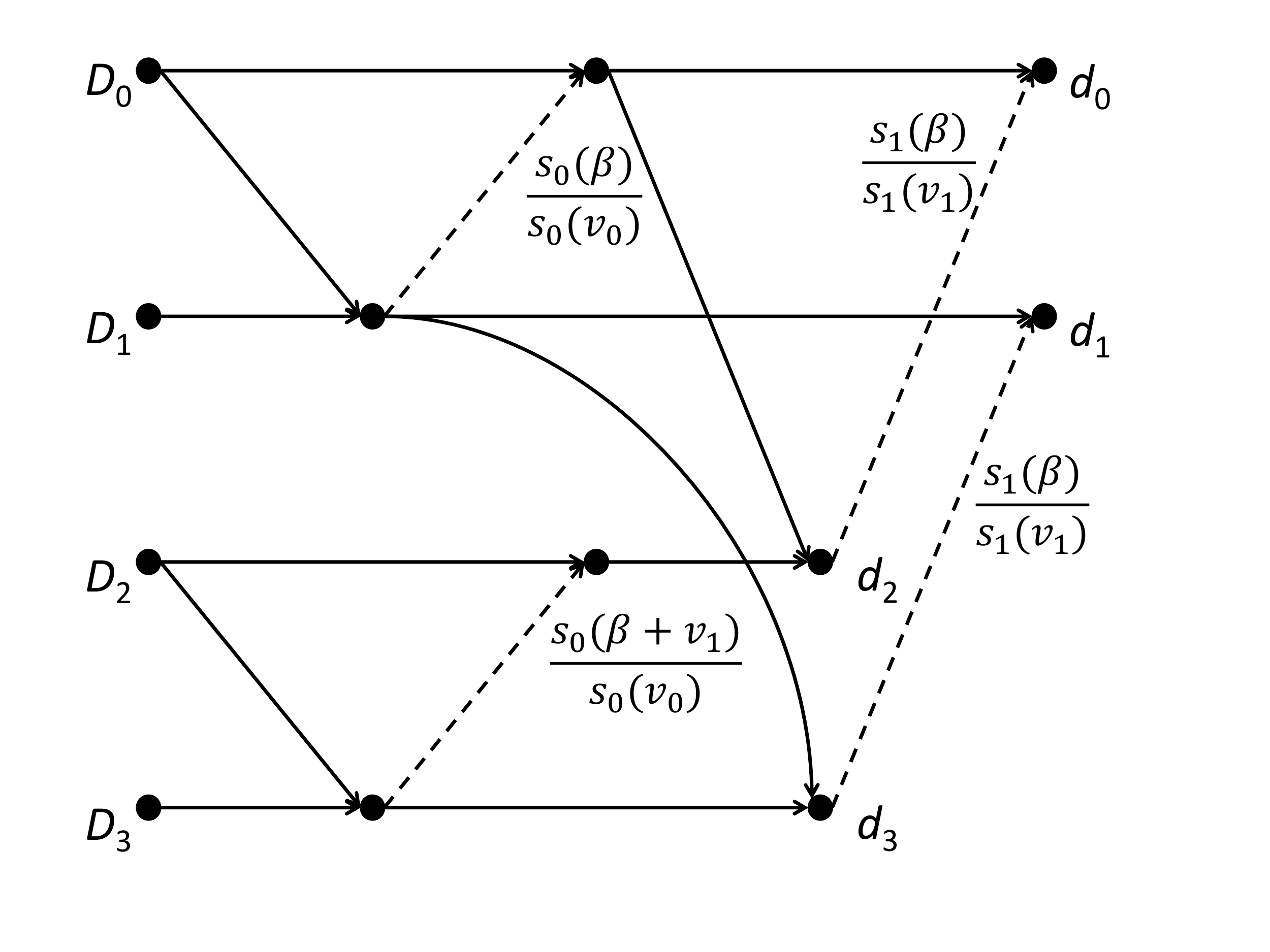}
   \label{fig:subfig2}
}
\caption{\label{fig:fig1}Data flow diagram of proposed $4$-poiont transform and its inversion.}
\end{figure}

For any polynomial $f(x)$ and a set $V$, let the notation $f(V)$ denote a set of evaluation values $f(V)=\{f(a)|\forall a\in V\}$. \cite{4319038} gave a recursive algorithm in $O(2^k\lg (2^k))$ to calculate $\bar D_{2^k}(V_k+\beta)$, where
\[
V_k+\beta =\{a+\beta|a \in V_k\}\mbox{ for any }\beta\in \mathbb{F}_{2^m}.
\]
In this section, we describe the algorithm~\cite{4319038} in another viewpoint, which helps us to develop encoding/decoding algorithm for RS codes.
 
The set of evaluation points can be divided into two individual subsets
\[
V_k+\beta=(V_{k-1}+\beta) \cup (V_{k-1}+v_{k-1}+\beta),
\]
where $(V_{k-1}+v_{k-1}+\beta)$ is the coset of $(V_{k-1}+\beta)$ by adding $v_{k-1}$. Accordingly, the set of polynomial evaluations can be divided into two subsets
\begin{equation}\label{eq:barD2^k}
\bar{D}_{2^k}(V_k+\beta)=\bar{D}_{2^k}(V_{k-1}+\beta) \cup \bar{D}_{2^k}(V_{k-1}+v_{k-1}+\beta).
\end{equation}
The algorithm relied on the following lemma.
\begin{lemma}\label{MultipointEvaluation}
Given $\gamma \in \mathbb{F}_{2^m}$ and a polynomial $\bar{D}_{2^k}(x)\in \mathbb{F}_{2^m}[x]/(x^{2^m}-x)$ in the basis $\mathbb{\bar{X}}$, we have
\begin{equation}\label{eqn_db000}
\begin{aligned}
&\bar{D}_{2^k}(a+\gamma)\\
=&\sum_{i=0}^{2^{k-1}-1}(\bar{d}_i+\frac{s_{k-1}(\gamma)}{s_{k-1}(v_{k-1})}\bar{d}_{i+2^{k-1}})\bar{X}_i(a+\gamma),
\end{aligned}
\end{equation}
for each $a\in V_{k-1}$.
\end{lemma}

Based on Lemma \ref{MultipointEvaluation}, the algorithm to compute \eqref{eq:barD2^k} is described below. By substituting $\gamma =\beta$ into \eqref{eqn_db000}, we obtain
\begin{equation}\label{eqn_db1} 
\begin{aligned}
&\bar{D}_{2^k}(a+\beta)\\
=&\sum_{i=0}^{2^{k-1}-1}(\bar{d}_i+\frac{s_{k-1}(\beta)}{s_{k-1}(v_{k-1})}\bar{d}_{i+2^{k-1}})\bar{X}_i(a+\beta)\\
=&\sum_{i=0}^{2^{k-1}-1}g_i^{(0)} \bar{X}_i(a+\beta)=D_{2^{k-1}}^{(0)}(a+\beta)\qquad \forall a\in V_{k-1},
\end{aligned}
\end{equation}
where each 
\begin{equation}\label{eqn_db11} 
g_i^{(0)}=\bar{d}_i+\frac{s_{k-1}(\beta)}{s_{k-1}(v_{k-1})}\bar{d}_{i+2^{k-1}}\qquad i=0,1,\dots ,2^{k-1}-1.
\end{equation}
This converts $\bar{D}_{2^k}(V_{k-1}+\beta)$ into $D_{2^{k-1}}^{(0)}(V_{k-1}+\beta)$. Furthermore, by substituting $\gamma =v_{k-1}+\beta$ into \eqref{eqn_db000}, we obtain \eqref{eqn_db2},
\begin{figure*}[b]
\hrulefill
\begin{equation}\label{eqn_db2} 
\begin{aligned}
&\bar{D}_{2^k}(a+v_{k-1}+\beta)\\
=&\sum_{i=0}^{2^{k-1}-1}(\bar{d}_i+\frac{s_{k-1}(v_{k-1}+\beta)}{s_{k-1}(v_{k-1})}\bar{d}_{i+2^{k-1}})\bar{X}_i(a+v_{k-1}+\beta)
\\
=&\sum_{i=0}^{2^{k-1}-1}(\bar{d}_i+\frac{s_{k-1}(\beta)}{s_{k-1}(v_{k-1})}\bar{d}_{i+2^{k-1}}+\bar{d}_{i+2^{k-1}})\bar{X}_i(a+v_{k-1}+\beta) \\
=&\sum_{i=0}^{2^{k-1}-1}(g_i^{(0)}+\bar{d}_{i+2^{k-1}})\bar{X}_i(a+v_{k-1}+\beta)\\
=&\sum_{i=0}^{2^{k-1}-1}g_i^{(1)}\bar{X}_i(a+v_{k-1}+\beta)=\bar{D}_{2^{k-1}}^{(1)}(a+v_{k-1}+\beta)\qquad \forall a\in V_{k-1},\\
\end{aligned}
\end{equation}
\end{figure*}
where each
\begin{equation}\label{eqn_db21} 
g_i^{(1)}=g_i^{(0)}+\bar{d}_{i+2^{k-1}}\qquad i=0,1,\dots ,2^{k-1}-1.
\end{equation}
This converts $\bar{D}_{2^k}(V_{k-1}+v_{k-1}+\beta)$ into $D_{2^{k-1}}^{(1)}(V_{k-1}+v_{k-1}+\beta)$.

From \eqref{eqn_db1} \eqref{eqn_db2}, the set of evaluation points \eqref{eq:barD2^k} can be expressed as
\begin{equation}\label{eq:Rec2}
\begin{aligned}
&\bar{D}_{2^k}(V_k+\beta)\\
=&\bar{D}_{2^{k-1}}^{(0)}(V_{k-1} +\beta) \cup \bar{D}_{2^{k-1}}^{(1)}(V_{k-1}+v_{k-1} +\beta).
\end{aligned}
\end{equation}

By comparing \eqref{eq:barD2^k} and \eqref{eq:Rec2}, the degrees of both polynomials are reduced one-half (the number of terms are reduced from $2^k$ to $2^{k-1}$). The complexity of obtaining both polynomials are discussed below. In \eqref{eqn_db1}, each coefficient $g_i$ takes an addition and a multiplication, except if  $s_{k-1}(\beta )=0$, then $g_i=\bar{d}_i$ without any arithmetic operations. However, we do not consider this exception here, because the reduction from those exceptions is limited. As $D_{2^{k-1}}^{(0)}(x)$ has $2^{k-1}$ coefficients, it takes a total of $2^{k-1}$ additions and $2^{k-1}$ multiplications to obtain them. In \eqref{eqn_db2}, calculating each coefficient $g_i+\bar{d}_{i+2^{k-1}}$ takes an addition, so it takes a total of $2^{k-1}$ additions to obtain the coefficients of $\bar{D}_{2^{k-1}}^{(1)}(x)$.

This procedure can be applied recursively to each set $\bar{D}_{2^{k-1}}^{(0)}(V_{k-1} +\beta)$ and $\bar{D}_{2^{k-1}}^{(1)}(V_{k-1}+v_{k-1} +\beta)$ until the size of each set is one. With the divide-and-conquer strategy, the additive complexity and the multiplicative complexity are respectively written as
\[
A(h)=2\times A(h/2)+h, \qquad M(h)=2\times M(h/2)+h/2,
\]
and the result is $A(h)=h\lg(h)$ and $M(h)=h/2\lg(h)$. Algorithm \ref{alg:FFT} depicts the details of the recursive approach, denoted as $\mathrm{FFT}_\mathbb{\bar{X}}(\bullet,k ,\beta)$.

The inverse FFT can be obtained by backtracking FFT given above. As opposite to \eqref{eq:Rec2}, the inverse transform get the coefficients of $\bar{D}_{2^{k-1}}^{(0)}(x)$ and $\bar{D}_{2^{k-1}}^{(1)}(x)$, and the objective is to find the coefficients of $\bar{D}_{2^k}(x)$. We reformulate \eqref{eqn_db21} and \eqref{eqn_db11} as
\begin{equation}\label{eqn_db21I} 
\begin{aligned}
&\bar{d}_{i+2^{k-1}}=g_i^{(0)}+g_i^{(1)},\\
&\bar{d}_i=g_i^{(0)}+\frac{s_{k-1}(\beta)}{s_{k-1}(v_{k-1})}\bar{d}_{i+2^{k-1}}\qquad i=0,1,\dots ,2^{k-1}-1.
\end{aligned}
\end{equation}
From \eqref{eqn_db21I}, we can compute the coefficients of $\bar{D}_{2^k}(x)$. The coefficients of $\bar{D}_{2^{k-1}}^{(0)}(x)$ and $\bar{D}_{2^{k-1}}^{(1)}(x)$ can be obtained by applying the inverse transform recursively. The details are shown in Algorithm \ref{alg:IFFT}. Note that $\mathrm{IFFT}_\mathbb{\bar{X}}(\bullet,k, \beta)$ denotes the inverse transform. Algorithms \ref{alg:FFT} and \ref{alg:IFFT} use the same notations such that one can follow them easily. It is clear  that both algorithms have the same number of arithmetic operations. Figure \ref{fig:fig1} showed an example of the proposed algorithm and its inversion. The input polynomial is defined as $\bar{D}(x)=\sum_{i=0}^{3}d_i\bar{X}_i(x)$, and the output is given by $D_i=\bar{D}(\omega_i+\beta)$, for $i=0,1,2,3$.

\section{Polynomial basis with monic polynomials and its operations}\label{sec:polynomialDivision}
In this section, we define an alternative version of the polynomial basis, and its algorithms to perform multiplications, formal derivatives, and divisions on the new basis. All these operations will be used in the coding algorithms. The alternative basis is defined as
\[
\mathbb{X}=\{X_0(x),X_1(x),\dots ,X_{2^m-1}(x)\}
\]
in $\mathbb{F}_{2^m}[x]/(x^{2^m}-x)$, where each $X_i(x)$ is given in \eqref{eq:barX_i(x)}. This implies that each $X_i(x)$ is a monic polynomial. For any $D_{h-1}(x)\in \mathbb{F}_{2^m}[x]/(x^{2^m}-x)$, the basis conversion between $\mathbb{X}$ and $\mathbb{\bar{X}}$ requires only $h$ multiplications/divisions:
\begin{equation}\label{eq:basischange}
D_{h-1}(x)=\sum_{i=0}^{h-1}d_i\cdot\bar{X}_i(x)=\sum_{i=0}^{h-1}\frac{d_i}{p_i}X_i(x).
\end{equation}
With the linear-time basis conversion, the multipoint evaluation in $\mathbb{\bar{X}}$ (Algorithm \ref{alg:FFT}) can also be applied on $\mathbb{X}$, and the complexity is unchanged.

To simplify the notations, in the rest of this paper, the polynomials are represented in $\mathbb{X}$. For $D_{2^k}(x)$ in $\mathbb{X}$, the evaluations at $V_k+\beta =\{\omega_i+\beta\}_{i=0}^{2^k-1}$ is denoted as
\begin{equation}
\begin{aligned}
&\mathrm{FFT}_\mathbb{X}(D_{2^k}, k, \beta)\\
=&(D_{2^k}(\omega_0+\beta),D_{2^k}(\omega_1+\beta),\dots ,D_{2^k}(\omega_{2^k-1}+\beta)),
\end{aligned}
\end{equation}
and the inversion is denoted as $\mathrm{IFFT}_\mathbb{X}(D_{2^k}, k, \beta)$. Based on Algorithm \ref{alg:FFT}, the transforms are defined as
\begin{equation}
\begin{aligned}
\mathrm{FFT}_\mathbb{X}(D_{2^k}, k, \beta)&=\mathrm{FFT}_\mathbb{\bar{X}}(D_{2^k}\otimes P_{2^k}, k, \beta),\\
\mathrm{IFFT}_\mathbb{X}(D_{2^k},k, \beta)&=\mathrm{IFFT}_\mathbb{\bar{X}}(D_{2^k}, k, \beta)\oslash P_{2^k},
\end{aligned}
\end{equation}

where $P_{2^k}=(p_0,p_1,\dots ,p_{2^k-1})$. The operation $\otimes$ is the pairwise multiplication on two vectors, and the operation $\oslash$ is the pairwise division. Since the multiplication and formal derivative in $\mathbb{X}$ are similar to those given in~\cite{4319038}, we summarize them in Appendix \ref{sec:operation2} for completeness. Next we present the algorithm for polynomial division that is essential for decoding of RS codes.

\subsection{Polynomial Division}\label{sec:division}
In this subsection, we proposed an $\mathcal{O}(h\lg (h))$ polynomial division in the basis $\mathbb{X}$. The proposed algorithm is based on Newton iteration approach that was used by the fast division algorithms in the standard basis~\citep{Gathen:2013} with $\mathcal{O}(h\lg (h))$, if $\mathcal{O}(h\lg (h))$ FFT exists. However, since our basis is different from the standard basis, some moderate modifications are required. 

As compared with the conventional fast division~\citep{Gathen:2013}, the proposed approach has two major differences. First, the conventional fast division shall reverse the coefficients of the divisor $B(x)$ upon performing the Newton iteration. However, in our basis $\mathbb{X}$,  the polynomial reversion cannot be applied. Thus, the proposed algorithm does not reverse the polynomials, and all operations are performed on the polynomials without reversions. Second, the proposed algorithm includes some specific multiplications that are not required in the conventional approach, such as $X_y(x)$ in \eqref{eq:A_mathbbX(x)} and $s_1(x)$ in \eqref{eq:lambda(i)(x)}. The objective of these multiplications are to align the results such that the desired polynomial can be extracted properly.

Let $\mathbb{Q}(A(x), i)$ denote the quotient of dividing $A(x)$ by $s_i(x)$, where $A(x)$ is in the basis $\mathbb{X}$ and $\deg (A(x))<2^{i+1}$. Precisely, for a polynomial of degree $h<2^{i+1}$, 
\begin{equation}
\begin{aligned}
A(x)&=\sum_{l=0}^{h-1}a_lX_l(x)=\sum_{l=0}^{2^i-1}a_lX_l(x)+\sum_{l=2^i}^{h-1}a_lX_l(x)\\
&=\sum_{l=0}^{2^i-1}a_lX_l(x)+s_i(x)\sum_{l=2^i}^{h-1}a_lX_{l-2^i}(x).
\end{aligned}
\end{equation}
The quotient of dividing $A(x)$ by $s_i(x)$ is then
\[
\mathbb{Q}(A(x), i)=\sum_{l=0}^{h-1-2^i}a_{l+2^i}X_l(x).
\]

In general, given a dividend $a(x)$ and a divisor $b(x)$, the division is to determine the quotient $Q(x)$ and the remainder $r(x)$ such that
\begin{equation}\label{eq:A_mathbbX2(x)}
a(x)=Q(x)\cdot b(x)+r(x),
\end{equation}
where $\deg(r(x))\leq \deg(b(x))-1$. Without loss of generality, we consider the case
\begin{equation}\label{eq:con}
\deg(a(x))> \deg(b(x))\geq 0.
\end{equation}

The proposed algorithm firstly finds out the quotient $Q(x)$, and then the remainder is calculated by 
\begin{equation}\label{eq:AmathbbX(x)2}
r(x)=a(x)-Q(x)\cdot b(x).
\end{equation}
In the following, we focus on the algorithm to determine $Q(x)$. 

Let
\begin{equation}\label{eq:y}
y=2^{D_\ell }-\deg(b(x))-1,
\end{equation}
where
\begin{equation}\label{eq:Dell}
D_\ell = \lceil\lg(\deg(a(x))+1)\rceil.
\end{equation}
To begin with, \eqref{eq:AmathbbX(x)2} is multiplied by $X_y(x)$ to obtain
\begin{equation}\label{eq:A_mathbbX(x)}
r(x)\cdot X_y(x)=a(x)\cdot X_y(x)-Q(x)\cdot b(x)\cdot X_y(x).
\end{equation}
To simplify the notations, let 
\begin{equation}\label{eq:A_mathbbX(x)3}
\begin{aligned}
R(x)&=r(x)\cdot X_y(x),\\
A(x)&=a(x)\cdot X_y(x),\\
B(x)&=b(x)\cdot X_y(x).
\end{aligned}
\end{equation}

Then we have
\begin{equation}\label{eq:A_mathbbX(x)-1}
 R(x)=A(x)-Q(x)\cdot B(x).
\end{equation}
Next we present a method to determine $Q(x)$ from \eqref{eq:A_mathbbX(x)-1}. 

Assume that there exists a polynomial $\Lambda(x)$ such that
\begin{equation}\label{eq:LambdamathbbX(x)}
\Lambda(x)\cdot s_1(x)\cdot B(x)=s_{D_a}(x)+H(x),
\end{equation}
where 
\begin{equation}\label{eq:LambdamathbbX(x)1}
\deg(H(x))\leq \deg(B(x))+1=2^{D_\ell},
\end{equation}
and
\begin{equation}\label{eq:D_a}
D_a=\lceil\lg(\deg(A(x))+1)\rceil.
\end{equation}
The algorithm to find out $\Lambda(x)$ will be addressed in Section~\ref{sec:determiningLambda}.  
Before determining $Q(x)$, we first present two  lemmas whose proofs are given in Appendix~\ref{lemma-proof}.
\begin{lemma}\label{lemmaDa}
\begin{equation}\label{eq:D_a=}
D_a=D_\ell +1.
\end{equation}
\end{lemma}
From \eqref{eq:LambdamathbbX(x)} and Lemma \ref{lemmaDa}, the degree of $\Lambda(x)$ is thus 
\begin{equation}\label{eq:degLambda}
\begin{aligned}
&\deg(\Lambda(x))\\
=& \deg (s_{D_a}(x))-\deg(B(x))-\deg (s_1(x))\\
=&2^{D_a}-(2^{D_\ell}-1)-2= 2^{D_\ell}-1.
\end{aligned}
\end{equation}
After obtaining $\Lambda(x)$, \eqref{eq:A_mathbbX(x)-1} is multiplied by $\Lambda(x)\cdot s_1(x)$ to obtain
\begin{equation}
\begin{aligned}
&R(x)\cdot \Lambda(x)\cdot s_1(x)\\
=& A(x)\cdot \Lambda(x)\cdot s_1(x)-Q(x)\cdot B(x)\cdot \Lambda(x)\cdot s_1(x)
\end{aligned}
\end{equation}
By \eqref{eq:LambdamathbbX(x)}, we have
\begin{equation*}
\begin{aligned}
&R(x)\cdot \Lambda(x)\cdot s_1(x)\\
=&A(x)\cdot \Lambda(x)\cdot s_1(x)-Q(x)\cdot (s_{D_a}(x)+H(x))
\end{aligned}
\end{equation*}
and then
\begin{equation}\label{eq:AmathbbX(x)}
\begin{aligned}
&Q(x)\cdot H(x)+R(x)\cdot \Lambda(x)\cdot s_1(x)\\
=&A(x)\cdot \Lambda(x)\cdot s_1(x)-Q(x)\cdot s_{D_a}(x).
\end{aligned}
\end{equation}

\begin{lemma}\label{lemmaQx}
The left-hand side of \eqref{eq:AmathbbX(x)} has  degree
\begin{equation}\label{eq:lemmaQx}
\deg(Q(x)\cdot H(x)+R(x)\cdot \Lambda(x)\cdot s_1(x)) \leq 2^{D_a}-1.
\end{equation}
\end{lemma}
In \eqref{eq:AmathbbX(x)}, $Q(x)\cdot s_{D_a}(x)$ is a polynomial where the coefficients of $Q(x)$ starts from $X_{2^{D_a}}(x)=s_{D_a}(x)$. By Lemma \ref{lemmaQx}, the degree of the left-hand side is no more than $2^{D_a}-1$. Thus, $A(x)\cdot \Lambda(x)\cdot s_1(x)$ has quotient $Q(x)$ starting on degree $2^{D_a}$, and hence the quotient can be obtained by
\begin{equation}\label{eq:AmathbbX(x)3}
Q(x)=\mathbb{Q}(A(x)\cdot \Lambda(x)\cdot s_1(x), D_a),
\end{equation}
In \eqref{eq:AmathbbX(x)3}, we have
\begin{equation}\label{eq:AmathbbX(x)4}
\deg(A(x)\cdot \Lambda(x)\cdot s_1(x))\leq 2^{D_a}+2^{D_\ell-1}+2=3\cdot 2^{D_\ell}+1.
\end{equation}

Algorithm \ref{alg:polydivision} shows the steps of the division algorithm. The complexity is analyzed below. In Step 1, as $\deg (A(x))=\deg (a(x))+\deg (y(x))<2^{D_\ell+1}$ and $\deg (B(x))=\deg (b(x))+\deg (y(x))=2^{D_\ell}-1$, the complexity is $\mathcal{O}(2^{D_\ell+1}\lg (2^{D_\ell+1}))=\mathcal{O}(2^{D_\ell} \lg (2^{D_\ell}))$. In Step 2, we will show that $\mathcal{O}(2^{D_\ell} \lg (2^{D_\ell}))$ suffice  in Section~\ref{sec:determiningLambda}. In Step 3, \eqref{eq:AmathbbX(x)4} shows that the complexity is $\mathcal{O}(2^{D_\ell} \lg (2^{D_\ell}))$. In Step 4, as the degrees of polynomials are less than $2^{D_\ell}$, the complexity is no more than $\mathcal{O}(2^{D_\ell} \lg (2^{D_\ell}))$. In summary, Algorithm \ref{alg:polydivision} has the complexity $\mathcal{O}(2^{D_\ell} \lg (2^{D_\ell}))=\mathcal{O}(\deg(a(x)) \lg (\deg(a(x))))$.

\begin{algorithm}[t]
\caption{\label{alg:polydivision} Polynomial divisions in $\mathbb{X}$}
\begin{algorithmic}[1]
\Require
A dividend $a(x)$ and a divisor $b(x)$, with $\deg (a(x))>\deg(b(x))\geq 0$
\Ensure
A quotient $Q(x)$ and a remainder $r(x)$, such that
\[
a(x)=Q(x)\cdot b(x)+r(x).
\]
\State Compute
\[
A(x)=a(x)\cdot X_y(x),
\]
\[
B(x)=b(x)\cdot X_y(x),
\]
where $y$ is defined as \eqref{eq:y}.
\State Find $\Lambda(x)$ such that \eqref{eq:LambdamathbbX(x)} holds.
\State Compute $Q(x)$ by \eqref{eq:AmathbbX(x)3}.
\State Compute $r(x)$ by \eqref{eq:AmathbbX(x)2}.
\State \Return $Q(x)$ and $r(x)$.
\end{algorithmic}
\end{algorithm}

\subsection{Determining $\Lambda(x)$ given in \eqref{eq:LambdamathbbX(x)}}\label{sec:determiningLambda}
Given $B(x)=\sum_{j=0}^{d_B} b_jX_j(x)$ with $b_{d_B}\neq 0$, this subsection presents a method to find out $\Lambda(x)$ in \eqref{eq:LambdamathbbX(x)}. Notice that $d_B=\deg(\Lambda(x))= 2^{D_\ell}-1$. The proposed method can be seen as a modified version of the division with Newton iterations~\cite{Gathen:2013}\cite{Mateer:2008}.

The method iteratively computes the coefficients of $\Lambda(x)$ from highest degree to lowest degree. For $i=0,1,\dots ,D_\ell$, the updated polynomial $\Lambda_i(x)$ of degree $2^i-1$ is calculated from $\Lambda_{i-1}(x)$. The initial polynomial is
\begin{equation}\label{eq:lambda_0(x)}
\Lambda_0(x)=b_{d_B}^{-1}.
\end{equation}

Let $B_{D_\ell}(x)=B(x)$, and
\begin{equation}\label{eq:B_i(x)2}
B_i(x)=\mathbb{Q}(B_{i+1}(x), i) \qquad i=0,1,\dots, D_\ell-1.
\end{equation}
\eqref{eq:B_i(x)2} can  be rewritten as
\begin{equation}\label{eq:B_i(x)3}
B_{i+1}(x)=B_i(x)\cdot s_i(x)+\bar{B}_i(x),
\end{equation}
where $\bar{B}_i(x)$, $\deg (\bar{B}_i)\leq 2^i-1$, is the residual. Clearly, $\deg (B_i(x))=2^i-1$. 

For $i=1,2,\ldots, D_\ell$, $\Lambda_i(x)$ is defined as
\begin{equation}\label{eq:lambda(i)'(x)}
\Lambda_i(x)=\mathbb{Q}((s_{i-1}(x))^2\cdot \bar{\Lambda}_i(x), i+1),
\end{equation}
where
\begin{equation}\label{eq:lambda(i)(x)}
\bar{\Lambda}_i(x)=(\Lambda_{i-1}(x))^2\cdot B_i(x)\cdot s_1(x).
\end{equation}
It can be verified that $\deg(\bar{\Lambda}_i(x))=2^{i+1}-1$ and $\deg(\Lambda_i(x))=2^i-1$ holds. The validity of  
$\Lambda(x)=\Lambda_{D_\ell}(x)$ is supported as follows, where all proofs are given in Appendix~\ref{lemma-proof}.

\begin{lemma}\label{lemma4}
$\Lambda_i(x)$ possesses the following equality:
\begin{equation}\label{eq:lambda(i)}
\Lambda_i(x)\cdot B_i(x)\cdot s_1(x)=s_{i+1}(x)+\bar{r}_i(x),
\end{equation}
with $\deg(\bar{r}_i(x))\leq 2^{i}$.
\end{lemma}
The following reformulation of  \eqref{eq:lambda(i)'(x)}, that contains no polynomial multiplications, can be used to determine the complexity of calculating \eqref{eq:lambda(i)'(x)}.
\begin{lemma}\label{lemma5}
\eqref{eq:lambda(i)'(x)} can be rewritten as
\begin{equation}\label{eq:lambda(i)'(x)5}
\Lambda_i(x)=\Lambda_i^{(1)}(x)+\mathbb{Q}(\Lambda_i^{(1)}(x), i-1)\cdot s_{i-1}(v_{i-1}),
\end{equation}
where
\[
\Lambda_i^{(1)}(x)=\mathbb{Q}(\bar{\Lambda}_i(x), i).
\]
\end{lemma}

Algorithm \ref{alg:newton} depicts the steps. The algorithm repeats performing \eqref{eq:lambda(i)(x)} and \eqref{eq:lambda(i)'(x)} (or \eqref{eq:lambda(i)'(x)5}) to obtain $\Lambda_{D_\ell}(x)$, which is the desired output $\Lambda(x)$. For the complexity, each iteration (lines 3-4) calculates \eqref{eq:lambda(i)(x)} and \eqref{eq:lambda(i)'(x)}. In \eqref{eq:lambda(i)(x)}, as $\deg (\Lambda_{i-1}(x))=2^{i-1}-1$, $\deg(B_i(x))=2^i-1$ and $\deg (s_i(x))=2$, the multiplications~\eqref{eq:lambda(i)(x)} requires $\mathcal{O}(2^i\lg (2^i))$. In \eqref{eq:lambda(i)'(x)}, Lemma \ref{lemma5} showed that the computation can be reduced to $\mathcal{O}(2^i)$ without polynomial multiplications. Thus, each iteration takes $\mathcal{O}(2^i\lg (2^i))$ operations, and the complexity for the loop (line 2-5) takes
\[
\sum_{i=1}^{D_\ell}\mathcal{O}(2^i\lg (2^i))=\mathcal{O}(2^{D_\ell}\lg (2^{D_\ell})).
\]

\begin{algorithm}[t]
\caption{\label{alg:newton} $\Lambda(x)$ computation}
\begin{algorithmic}[1]
\Require
A polynomial $B(x)$
\Ensure
A polynomial $\Lambda(x)$ such that \eqref{eq:LambdamathbbX(x)} holds, where $\deg(\Lambda(x))=\deg (B(x))=2^{D_\ell}-1$.
\State Let $\Lambda_0(x)=b_{d_B}^{-1}$.
\For{$i=1,2,\dots ,D_\ell$} 
\State Compute \eqref{eq:lambda(i)(x)}.
\State Compute \eqref{eq:lambda(i)'(x)} (or \eqref{eq:lambda(i)'(x)5}, equivalently).
\EndFor
\State \Return $\Lambda_{D_\ell}(x)$.
\end{algorithmic}
\end{algorithm}

\section{Extended Euclidean Algorithm based on Half-GCD Approach}\label{sec:Euclidean}
\begin{algorithm}[htp]
\caption{\label{alg:HalfGCD} Half-GCD algorithm}
\begin{algorithmic}[1]
\Require
$\mathrm{HGCD}(a(x), b(x), g)$, where $a(x), b(x)\in \mathbb{F}_{2^m}[x]/x^{2^m}-x$ in basis $\mathbb{X}$, and $\deg(b(x))\le\deg(a(x))$, $2^{g-1}\leq\deg(a(x))\leq 2^g-1$
\Ensure Two matrices $(Z,M)$ given in \eqref{eq:ZM}
\If {$\deg(b(x))<2^{g-1}$}
\Return
\begin{equation}\label{eq:Eucint}
Z=\begin{bmatrix}
a(x)\\ 
b(x)
\end{bmatrix},\qquad
M=
\begin{bmatrix}
1 & 0\\ 
0 & 1
\end{bmatrix}.
\end{equation}
\EndIf
\State $(Z_\mathrm{H}, M_\mathrm{H})\leftarrow\mathrm{HGCD}(a_\mathrm{H}(x), b_\mathrm{H}(x), g-1)$
\State Compute
\begin{equation}\label{eq:EucComp1}
\begin{bmatrix}
z_\mathrm{M0}(x)\\ 
z_\mathrm{M1}(x)
\end{bmatrix}=Z_\mathrm{H}\cdot s_{g-1}(x)+
M_\mathrm{H}
\begin{bmatrix}
a_\mathrm{L}(x)\\ 
b_\mathrm{L}(x)
\end{bmatrix}.
\end{equation}

\If {$\deg(z_\mathrm{Ml}(x))\leq 2^{g-1}-1$} 
\Return 
\[
(Z_\mathrm{M}=
\begin{bmatrix}
z_\mathrm{M0}(x)\\ 
z_\mathrm{Ml}(x)
\end{bmatrix}, M_\mathrm{H}).
\]
\EndIf
\State $z_\mathrm{M0}(x)$ is divided by $z_\mathrm{M1}(x)$ to get
\begin{equation}\label{eq:dividEuclid}
z_\mathrm{M0}(x)=q_\mathrm{M}(x)\cdot z_\mathrm{M1}(x)+r_\mathrm{M}(x)
\end{equation}
with $\deg(r_\mathrm{M}(x))<\deg(z_\mathrm{M1}(x))\leq 2^{g-1}+2^{g-2}-1$.
\State
$z_\mathrm{M1}(x)$ and $r_\mathrm{M}(x)$ are divided into three polynomials, denoted as
\begin{equation*}
\begin{aligned}
&z_\mathrm{M1}(x)\\
=&z_\mathrm{M1LL}(x)+s_{g-2}(x)z_\mathrm{M1LH}(x)+s_{g-1}(x)z_\mathrm{M1H}(x),\\
&r_\mathrm{M}(x)\\
=&r_\mathrm{MLL}(x)+s_{g-2}(x)r_\mathrm{MLH}(x)+s_{g-1}(x)r_\mathrm{MH}(x).
\end{aligned}
\end{equation*}
Compute
\begin{equation*}
\begin{aligned}
&z_\mathrm{M1M}(x)\\
=&z_\mathrm{M1LH}(x)+(s_{g-2}(x)+s_{g-2}(v_{g-2}))z_\mathrm{M1H}(x),\\
&r_\mathrm{MM}(x)\\
=&r_\mathrm{MLH}(x)+(s_{g-2}(x)+s_{g-2}(v_{g-2}))r_\mathrm{MH}(x).
\end{aligned}
\end{equation*}

\State $(Y_\mathrm{M},M_\mathrm{M})\leftarrow \mathrm{HGCD}(z_\mathrm{M1M}(x), r_\mathrm{MM}(x), g-1)$
\State \Return $(Z_\mathrm{R}, M_\mathrm{R})$, where
\begin{equation}\label{eq:Zresult}
\begin{aligned}
M_\mathrm{R}&=M_\mathrm{M}
\begin{bmatrix}
0 & 1\\ 
1 & -q_\mathrm{M}(x)
\end{bmatrix}
M_\mathrm{H},\\
Z_\mathrm{R}&=
Y_\mathrm{M}\cdot s_{g-2}(x)+
M_\mathrm{M}
\begin{bmatrix}
z_\mathrm{M1LL}(x)\\ 
r_\mathrm{MLL}(x)
\end{bmatrix}.
\end{aligned}
\end{equation}
\end{algorithmic}
\end{algorithm}

This section introduces the extended Euclidean algorithm that will be used in the decoding of RS codes. Given two polynomials $a(x)=r_{-1}(x)$, $b(x)=r_0(x)$, and 
\begin{equation}\label{eq:condEuc}
\deg(b(x))\le \deg(a(x))< 2^g,
\end{equation}
Euclidean algorithm is a procedure to recursively divide $r_{k-2}(x)$ by $r_{k-1}(x)$ to get
\[
r_{k-2}(x)=q_k(x)\cdot r_{k-1}(x)+r_k(x),
\]
with $\deg(r_k)<\deg(r_{k-1})$. The procedure stops at $r_N(x)=0$, and $r_{N-1}(x)$ is the greatest common divisor (gcd) of $a(x)$ and $b(x)$. An extension version, namely extended Euclidean algorithm, calculates $r_k(x)$ with a pair of polynomials $(u_k(x), v_k(x))$ in each iteration such that
\[
a(x)\cdot u_k(x)+b(x)\cdot v_k(x)=r_k(x).
\]

The $(k-1)$-th step of extended Euclidean algorithm can be expressed as a matrix form
\begin{equation}
\begin{bmatrix}
r_{k-2}(x)\\ 
r_{k-1}(x)
\end{bmatrix}=
\begin{bmatrix}
u_{k-2}(x) & v_{k-2}(x)\\ 
u_{k-1}(x) & v_{k-1}(x)
\end{bmatrix}\cdot
\begin{bmatrix}
a(x)\\ 
b(x)
\end{bmatrix}.
\end{equation}
The next step is shown as
\begin{equation}\label{eq:next}
\begin{bmatrix}
r_{k-1}(x)\\ 
r_{k}(x)
\end{bmatrix}=
\begin{bmatrix}
0 & 1\\ 
1 & -q_k(x)
\end{bmatrix}\cdot
\begin{bmatrix}
u_{k-2}(x) & v_{k-2}(x)\\ 
u_{k-1}(x) & v_{k-1}(x)
\end{bmatrix}\cdot
\begin{bmatrix}
a(x)\\ 
b(x)
\end{bmatrix}.
\end{equation}

The half-GCD algorithm~\cite{575320}\cite{Gathen:2013} calculates the temporal result of extended Euclidean algorithm at $s$-th step such that
\begin{equation}\label{eq:ecuoutput}
\deg(r_s(x))\leq 2^{g-1}-1.
\end{equation}

In this section, we present a half-GCD algorithm in basis $\mathbb{X}$. This approach will be performed to solve the error locator polynomial (see \eqref{eq:keyequation3}) in the decoding procedure of RS codes.

For polynomials in the monomial basis, there exist fast approaches in $\mathcal{O}(M(h)\lg(h))$ operations, where $M(h)$ denotes the complexity of multiplying two polynomials of degrees $h/2$ (see \cite[Algorithm 11.6]{Gathen:2013} or \cite[Figure 8.3]{Mateer:2008}). The idea comes from an observation that, the quotient $q_k(x)$ in \eqref{eq:next} is determined by the upper degree part of $r_{k-1}(x)$ and $r_k(x)$, and the lower degree part of $r_{k-1}(x)$ and $r_k(x)$ are not necessary. Fortunately, this observation is also applicable to our basis $\mathbb{X}$.

From the observation, we partition the inputs $a(x)$ (and $b(x)$) into several portions, so that the procedure can be applied on the portions of higher degrees. For the algorithms on monomial basis, it is simple to make such partitions. For basis $\mathbb{X}$, we have to choose partition points at  degrees $X_{2^{g-2}}(x)$ and $X_{2^{g-1}}(x)$. Precisely, $a(x)$ is divided into three polynomials $a_\mathrm{LL}(x)$, $a_\mathrm{LH}(x)$ and $a_\mathrm{H}(x)$ at $s_{g-2}(x)$ and $s_{g-1}(x)$, respectively. The representation is given by
\begin{equation}\label{eq:axdivid2}
\begin{aligned}
a(x)=&a_\mathrm{L}(x)+s_{g-1}(x)a_\mathrm{H}(x)\\
=&a_\mathrm{LL}(x)+s_{g-2}(x)a_\mathrm{LH}(x)+s_{g-1}(x)a_\mathrm{H}(x).
\end{aligned}
\end{equation}
Similarly, $b(x)$ is partitioned in the same manner:
\begin{equation}\label{eq:axdivide3}
\begin{aligned}
b(x)=&b_\mathrm{L}(x)+s_{g-1}(x)b_\mathrm{H}(x)\\
=&b_\mathrm{LL}(x)+s_{g-2}(x)b_\mathrm{LH}(x)+s_{g-1}(x)b_\mathrm{H}(x).
\end{aligned}
\end{equation}
Algorithm \ref{alg:HalfGCD} depicts the proposed algorithm $\mathrm{HGCD}(a(x), b(x), g)$, with $\deg(b(x))\le\deg(a(x))$ and $2^{g-1}\leq\deg(a(x))\leq 2^g-1$. The algorithm outputs two matrices
\begin{equation}\label{eq:ZM}
Z=\begin{bmatrix}
z_0(x)\\ 
z_1(x)
\end{bmatrix}\mbox{ and } 
M=\begin{bmatrix}
m_{00}(x) & m_{01}(x)\\ 
m_{10}(x) & m_{11}(x)
\end{bmatrix}
\end{equation}
such that
\begin{enumerate}
\item
\begin{equation}\label{eq:ZM1}
Z=M\cdot \begin{bmatrix}
a(x)\\ 
b(x)
\end{bmatrix};
\end{equation}
\item 
\begin{equation}\label{eq:cond1}
\begin{aligned}
\deg(z_0(x))&\geq 2^{g-1},\\
\deg(z_1(x))&\leq 2^{g-1}-1;
\end{aligned}
\end{equation}
\item 
\begin{equation}\label{eq:cond2}
\begin{aligned}
\deg(m_{11}(x))&\leq\deg(a(x))-\deg(z_0(x));
\end{aligned}
\end{equation}
\item 
\begin{equation}\label{eq:cond3}
\begin{aligned}
\deg(m_{i0}(x))&\leq\deg(m_{i1}(x)),\\
\deg(m_{0i}(x))&\leq\deg(m_{1i}(x)),\qquad i=0,1.\\
\end{aligned}
\end{equation}
\end{enumerate}
Before proving the validity of  Algorithm~\ref{alg:HalfGCD}, we give the following Lemmas whose proofs are given in Appendix~\ref{lemma-proof}.
\begin{lemma}\label{lemma6.1}
Algorithm~\ref{alg:HalfGCD} always outputs $Z$ and $M$ given in \eqref{eq:ZM} that satisfy~\eqref{eq:ZM1}.
\end{lemma}
\begin{lemma}\label{lemma6.2}
The recursive calls in $\mathrm{HGCD}(a(x), b(x), g)$ meet the requirements $\deg(b(x))<\deg(a(x))$ and $2^{g-1}\leq\deg(a(x))\leq 2^g-1$.
\end{lemma}
\begin{lemma}\label{lemma6.5}
Algorithm~\ref{alg:HalfGCD} always outputs $Z$ and $M$ given in \eqref{eq:ZM} that satisfies \eqref{eq:cond1}.
\end{lemma}
\begin{lemma}\label{lemma6.55}
Algorithm~\ref{alg:HalfGCD}  always outputs $Z$ and $M$ given in \eqref{eq:ZM} that satisfy \eqref{eq:cond2} and \eqref{eq:cond3}.
\end{lemma}

By the above Lemmas, we have
\begin{theorem}\label{theorem2}
Algorithm~\ref{alg:HalfGCD} is valid. That is, Algorithm~\ref{alg:HalfGCD} always outputs $Z$ and $M$ given in \eqref{eq:ZM} that satisfy the above four conditions. 
\end{theorem}

We determine the computational complexity as follows. The algorithm complexity is denoted as $T(h)$ of polynomial degrees $h=2^g$. In step 3 and step 9, the algorithm shall call the routine twice, and it takes $2\cdot T(h/2)$. line 7 is the polynomial division, and this requires $\mathcal{O}(h\lg(h))$ by using the fast division approach in Sec. \ref{sec:polynomialDivision}. Line 4 and line 10 have polynomial additions and polynomials multiplications. As those polynomials have degrees less than $h$, the complexity is $\mathcal{O}(h\lg (h))$ by the results given in Appendix~\ref{sec:operation2}. In summary, the overall complexity is
\[
T(h)=2T(h/2)+\mathcal{O}(h\lg(h)),\mbox{ and } T(h)=\mathcal{O}(h\lg^2(h)).
\]

\section{Reed-Solomon encoding algorithm}\label{sec:Encoding}
This section introduces an $\mathcal{O}(n\lg (n-k))$ encoding algorithm for $(n=2^m,k)$ RS codes over $\mathbb{F}_{2^m}$, with $T=2^t=n-k$ a power of two. There exist two viewpoints for the constructions of RS codes, termed as the polynomial evaluation approach and the generator polynomial approach. For the polynomial evaluation approach, the message is interpreted as a polynomial $\mathbf{u}(x)\in \mathbb{F}_{2^m}[x]/(x^{2^m}-x)$ of degree less than $k$. The codeword $\mathbf{v}=(v_0,v_1,\dots ,v_{n-1})$ is defined as the evaluations of $\mathbf{u}(x)$ at $n$ distinct points.

Assume $\mathbf{u}(x)$ is in the basis $\mathbb{\bar{X}}$, and thus $\mathbf{u}(x)=\sum_{i=0}^{k-1}u_i \bar{X}_i(x)$. The vector of coefficients is denoted as
\begin{equation}\label{eq:u}
\mathbf{u}=(u_0,u_1,\dots ,u_{k-1},\underbrace{\omega_0,\omega_0,\dots ,\omega_0}_{T}),
\end{equation}
with $T$ $\omega_0$s in the high degree part.  Then the codeword can be computed via Algorithm~\ref{alg:FFT}:
\begin{equation}\label{eq:encoding0}
\mathbf{v}=\mathrm{FFT}_\mathbb{\bar{X}}(\mathbf{u},m,\omega_0).
\end{equation}

However, \eqref{eq:encoding0} requires $\mathcal{O}(n\lg (n))$ operations, and the generated codeword is not systematic. In the following, another formula with complexity $\mathcal{O}(n\lg (n-k))$ is given, and the generated codeword is systematic. The inversion of \eqref{eq:encoding0} is given by
\begin{equation}\label{eq:encoding1}
\mathbf{u}=\mathrm{IFFT}_\mathbb{\bar{X}}(\mathbf{v},m,\omega_0).
\end{equation}
Note that, in \eqref{eq:encoding1}, $\mathbf{u}$ has $T$ $\omega_0$s in the high degree part (see \eqref{eq:u}). To begin with, $\mathbf{v}$ is divided into a number of sub-vectors
\begin{equation}\label{eq:mathbfv}
\mathbf{v}=(\mathbf{v}_0,\mathbf{v}_1,\dots ,\mathbf{v}_{n/T-1}),
\end{equation}
where each $\mathbf{v}_i$ has $T$ elements defined as
\[
\mathbf{v}_i=(v_{i\cdot T}, v_{1+i\cdot T},\dots ,v_{T-1+i\cdot T})\qquad i=0,1,\dots , n/T-1.
\]
Those sub-vectors can be proved to possess the  equality given in the following lemma, whose proof is given in Appendix~\ref{lemma-proof}.

\begin{lemma}\label{lemma7}
The following equality is hold: 
\begin{equation}\label{eq:encoding2}
\begin{aligned}
\omega_0=&\mathrm{IFFT}_\mathbb{\bar{X}}(\mathbf{v}_0, t,\omega_0)+\mathrm{IFFT}_\mathbb{\bar{X}}(\mathbf{v}_1, t,\omega_T)+\dots \\
&+\mathrm{IFFT}_\mathbb{\bar{X}}(\mathbf{v}_{n/T-1}, t,\omega_k),
\end{aligned}
\end{equation}
where $+$ is the addition for vectors.
\end{lemma}

\eqref{eq:encoding2} plays the core transform of the proposed algorithm. Assume $\mathbf{v}_0$ includes the parity symbols, and others $\{\mathbf{v}_i\}_{i=1}^{n/T-1}$ are the message symbols. From \eqref{eq:encoding2}, the parity is computed via
\begin{equation}\label{eq:encoding3}
\begin{aligned}
\mathbf{v}'_0=&\mathrm{IFFT}_\mathbb{\bar{X}}(\mathbf{v}_1, t,\omega_T)+\mathrm{IFFT}_\mathbb{\bar{X}}(\mathbf{v}_2, t,\omega_{2T})+\dots \\
&+\mathrm{IFFT}_\mathbb{\bar{X}}(\mathbf{v}_{n/T-1}, t,\omega_k),\\
\mathbf{v}_0=&\mathrm{FFT}_\mathbb{\bar{X}}(\mathbf{v}'_0, t, \omega_0).
\end{aligned}
\end{equation}

This algorithm requires a $T$-point FFT and $(n/T-1)$ times of $T$-point IFFT. Hence, the complexity of the encoding algorithm is
\[
\mathcal{O}(T\lg(T))+(n/T-1)\mathcal{O}(T\lg(T))=\mathcal{O}(n\lg(n-k)).
\]

\section{Reed-Solomon decoding algorithm}\label{sec:Decoding}
This section shows a decoding algorithm for $(n=2^m,k)$ RS codes over $\mathbb{F}_{2^m}$, where the codeword $\mathbf{v}=(v_0,\dots ,v_{n-1})=(\mathbf{u}(\omega_0),\dots ,\mathbf{u}(\omega_{n-1}))$ is generated by Section~\ref{sec:Encoding}. The proposed algorithm follows the syndrome-based decoding process. Let $\mathbf{r}=(r_0,r_1,\ldots,r_{n-1})=\mathbf{v}+\mathbf{e}$ denote the received vector with error pattern $\mathbf{e}=(e_0,e_1,\ldots,e_{n-1})$. Hence,\begin{equation}\label{eq:ri}
r_i=\mathbf{u}(\omega_i)+e_i.
\end{equation}
If $e_i\neq 0$, $r_i$ is an erroneous symbol. Suppose $\mathbf{e}$ contains $v\leq (n-k)/2=T/2$ non-zero symbols. Let 
\begin{equation}\label{eq:E}
E=\{\omega_i\in \mathbf{F}_{2^m}|e_i\neq 0\}
\end{equation}
denote the set of $\omega_i$ corresponding to locations of errors. Then, error-locator polynomial is defined as 
\begin{equation}\label{eq:lambda}
\lambda(x)=\prod_{\omega_i\in E} (x-\omega_i).
\end{equation}
Let $\mathbf{\bar{r}}(x)$ denote a polynomial of degree less than $2^m$, with $\mathbf{\bar{r}}(\omega_i)=r_i,\forall \omega_i\in \mathbb{F}_{2^m}$. It is clear to see that
\[
\mathbf{\bar{r}}(\omega_i)\cdot \lambda(\omega_i)=\left\{\begin{array}{ll}
0 & \text{if } \omega_i\in E;\\ 
r_i\cdot \lambda(\omega_i)& \text{if } \omega_i\in \mathbb{F}_{2^m}\setminus E.
\end{array}\right.
\]
The above formula leads to 
\begin{equation}\label{eq:mathu}
\begin{aligned}
&\mathbf{u}(\omega_i)\cdot \lambda(\omega_i)=\mathbf{\bar{r}}(\omega_i)\cdot \lambda(\omega_i)\\
\Rightarrow &\mathbf{u}(x)\cdot \lambda(x)=\mathbf{\bar{r}}(x)\cdot \lambda(x)\pmod {x-\omega_i}\qquad \forall \omega_i\in \mathbb{F}_{2^m}.
\end{aligned}
\end{equation}
Due to
\[
\prod_{i=0}^{2^m-1}(x-\omega_i)=x^{2^m}-x=s_m(x),
\]
\eqref{eq:mathu} implies that
\begin{equation}\label{eq:keyequation}
\begin{aligned}
&\mathbf{u}(x)\cdot \lambda(x)\equiv\mathbf{\bar{r}}(x)\cdot \lambda(x) \pmod {s_m(x)}\\
\Rightarrow &\mathbf{u}(x)\cdot \lambda(x)=\mathbf{\bar{r}}(x)\cdot \lambda(x)+\mathbf{q}(x)\cdot s_m(x),
\end{aligned}
\end{equation}
with $\deg(\mathbf{q}(x))<v\leq T/2$. Given $\mathbf{\bar{r}}(x)$, \eqref{eq:keyequation} is the key equation~\cite{21269}\cite{Gao02anew} to find out $\lambda(x)$, by applying the Euclidean algorithm on $s_m(x)$ and $\mathbf{\bar{r}}(x)$. However, though \eqref{eq:keyequation} is similar to the key equation of the syndrome decoding, $\mathbf{\bar{r}}(x)$ is not the syndrome polynomial. To obtain the syndrome decoding, the new key formula is the quotients of dividing $\lambda(x)$ and $s_m(x)$ by $X_k(x)$.

In this case, $\mathbf{\bar{r}}(x)$ is divided into two parts
\begin{equation}\label{eq:mathbfbarr(x)}
\mathbf{\bar{r}}(x)=\mathbf{\bar{r}}_0(x)+X_k(x)\mathbf{s}(x),
\end{equation}
where $\mathbf{\bar{r}}_0(x)$ denotes the residual. Notably, if no error occurs, $\mathbf{\bar{r}}(x)=\mathbf{u}(x)$ of degree less than $k$, and hence $\mathbf{s}(x)=0$. Thus we can take $\mathbf{s}(x)$ as the syndrome polynomial. 

For $s_m(x)$, the polynomial is recursively decomposed by \eqref{eq:w_j2(x)-1} to obtain \eqref{eq:x^2^m0}.
\begin{figure*}[b]
\hrulefill
\begin{equation}\label{eq:x^2^m0}
\begin{aligned}
 s_m(x)=&s_{m-1}(x)\left(s_{m-1}(v_{m-1})+s_{m-1}(x)\right)\\
=&s_{m-1}(x)\left(s_{m-1}(v_{m-1})+s_{m-2}(x)\left(s_{m-2}(v_{m-2})+s_{m-2}(x)\right)\right)\\
=&\cdots \\
=&s_{m-1}(x)s_{m-1}(v_{m-1})+s_{m-1}(x)s_{m-2}(x)s_{m-2}(v_{m-2})+\dots \\
&+s_{m-1}(x)s_{m-2}(x)\dots s_{t}(x)s_{t}(v_t)+s_{m-1}(x)s_{m-2}(x)\dots s_{t}(x)s_{t}(x)\\
=&X_{2^m-2^{m-1}}(x)s_{m-1}(v_{m-1})+X_{2^m-2^{m-2}}(x)s_{m-2}(v_{m-2})+\dots +X_{2^m-2^t}(x)s_{t}(v_t)+X_{2^m-2^t}(x)s_{t}(x).
\end{aligned}
\end{equation}
\end{figure*}

In \eqref{eq:x^2^m0}, the degree of each term is less than $k=n-T=2^m-2^t$, except for the last term $X_{2^m-2^t}(x)s_t(x)$. Thus, the quotient of dividing $s_m(x)$ by $X_k(x)$ would be $s_t(x)$.

Based on above results, the new key formula is
\begin{equation}\label{eq:keyequation3}
\begin{aligned}
\mathbf{z}_0(x)= \mathbf{s}(x) \lambda(x)+\mathbf{q}(x) s_t(x),
\end{aligned}
\end{equation}
with $\deg(\mathbf{z}_0(x))\leq T/2$. \eqref{eq:keyequation3} is the key equation to find the error locator polynomial. 

To find $\lambda(x)$, extended Euclidean algorithm is applied on $s_t(x)$ and $\mathbf{s}(x)$. The extended Euclidean algorithm stops when the remainder has degree less than $T/2$. After obtaining $\lambda(x)$, the next step is to find out the locations of errors $E$ defined in~\eqref{eq:E}, that is the set of roots of $\lambda(x)$.

After obtaining $E$, the final step is to calculate the error values. The formal derivative of \eqref{eq:keyequation} is 
\begin{equation}\label{eq:KEFormalDerivative}
\begin{aligned}
&\mathbf{u}'(x)\cdot \lambda(x)+\mathbf{u}(x)\cdot \lambda'(x)\\
=&\mathbf{\bar{r}}'(x)\cdot \lambda(x)+\mathbf{\bar{r}}(x)\cdot \lambda'(x)+\mathbf{q}'(x)\cdot s_m(x)+\mathbf{q}(x).
\end{aligned}
\end{equation}
By substituting $\omega_i\in E$ into \eqref{eq:KEFormalDerivative}, the error value is given by
\begin{equation}\label{eq:errorValue}
\begin{aligned}
&\mathbf{u}(\omega_i)\cdot \lambda'(\omega_i)=\mathbf{\bar{r}}(\omega_i)\cdot \lambda'(\omega_i)+\mathbf{q}(\omega_i)\\
\Rightarrow &\mathbf{u}(\omega_i)-\mathbf{\bar{r}}(\omega_i)=\frac{\mathbf{q}(\omega_i)}{\lambda'(\omega_i)},
\qquad \forall \omega_i\in E.
\end{aligned}
\end{equation}
Notice that \eqref{eq:errorValue} uses $\mathbf{q}(x)$ to compute the error values, rather than $\mathbf{z}_0(x)$ used in Forney's formula. In summary, the decoding algorithm consists of four steps:
\begin{enumerate}
\item Calculate syndrome polynomial $\mathbf{s}(x)$.
\item Determine the error-locator polynomial $\lambda(x)$ from \eqref{eq:keyequation3} by extended Euclidean algorithm.
\item Find the error locations $E$.
\item Calculate the error values via \eqref{eq:errorValue}.
\end{enumerate}
The details of each step is described below. In the first step, $\mathbf{s}(x)$ is the high degree part of applying IFFT on the received codeword $\mathbf{r}$. However, since the high degree part is required only, we follow the same idea of the encoding formula~\eqref{eq:encoding2}. In particular, the received codeword is divided into several individual parts $\mathbf{r}=(\mathbf{r}_0,\mathbf{r}_1,\dots ,\mathbf{r}_{n/T-1})$, where each $\mathbf{r}_i$ has $T=2^t=n-k$ elements. Then the syndrome polynomial is calculated by
\begin{align*}
\mathbf{s}=&\mathrm{IFFT}_\mathbb{X}(\mathbf{r}_0, t, \omega_0)+\mathrm{IFFT}_\mathbb{X}(\mathbf{r}_1, t, \omega_T)+\dots \\
&+\mathrm{IFFT}_\mathbb{X}(\mathbf{r}_{n/T-1}, t, \omega_k).
\end{align*}

In the second step, the fast Euclidean algorithm~(Algorithm \ref{alg:HalfGCD}) is applied on $s_t(x)$ and $\mathbf{s}(x)$. Upon performing the Euclidean algorithm, we go a step by dividing $s_t(x)$ with $\mathbf{s}(x)$, resulting in 
\[
s_t(x) = q_\mathrm{t}(x)\cdot \mathbf{s}(x)+ r_\mathrm{t}(x).
\]
Then call Algorithm \ref{alg:HalfGCD} with inputs $\mathbf{s}(x)$ and $r_\mathrm{t}(x)$ to obtain
\[
(\begin{bmatrix}
z_0(x)\\ 
z_1(x)
\end{bmatrix}, \begin{bmatrix}
u_0(x) & v_0(x)\\ 
u_1(x) & v_1(x)
\end{bmatrix})\leftarrow \mathrm{HGCD}(\mathbf{s}(x), r_\mathrm{t}(x), T).
\]
Then we have
\begin{equation}
\begin{aligned}
&z_1(x)=u_1(x)\mathbf{s}(x)+v_1(x)r_\mathrm{t}(x)\\
\Rightarrow &z_1(x)=u_1(x)\mathbf{s}(x)+v_1(x)(s_t(x) - q_\mathrm{t}(x)\cdot \mathbf{s}(x))\\
\Rightarrow &z_1(x)=v_1(x)s_t(x)+(u_1(x)-v_1(x)q_\mathrm{t}(x))\mathbf{s}(x),
\end{aligned}
\end{equation}
and thus the error locator polynomial is given by 
\[
\lambda(x)=u_1(x)-v_1(x)q_\mathrm{t}(x). 
\]
In the third step, the roots of $\lambda(x)$ can be searched via FFTs. The transform 
\begin{equation}\label{eq:RootFinding}
\mathrm{FFT}_\mathbb{X}(\lambda,T, \omega_{i\cdot T})
\end{equation}
is to evaluate $\lambda(x)$ at $V_t+\omega_{i\cdot T}$. If the result vector contains zeros, then $\lambda(x)$ has some roots at the corresponding points. \eqref{eq:RootFinding} is performed at $i=0,1,\dots ,n/T-1$ to search the roots in $\mathbb{F}_{2^m}$. Notably, if $\deg(\lambda(x))$ is larger than the number of found roots, the decoding procedure shall be terminated. This situation occurs when the number of errors exceeds $T/2$.

In the final step, we compute $\mathrm{FFT}_\mathbb{X}(\mathbf{q}, T, \omega_{i\cdot T})$ and $\mathrm{FFT}_\mathbb{X}(\lambda', T, \omega_{i\cdot T})$ (computing $\lambda'(x)$ is given in Appendix \ref{sec:operation2}), for $i=0,1,\dots ,n/T-1$. Then the error values are calculated via \eqref{eq:errorValue}. 

To determine the computational complexity, the first step requires $(n/T)$ times of $T$-point IFFT such that the complexity is $n/T\cdot \mathcal{O}(T\lg(T))=\mathcal{O}(n\lg(n-k))$. The second step takes $\mathcal{O}((n-k)\lg^2(n-k))$ operations. The third step requires $(n/T)$ times of $T$-point FFT, and thus the complexity is $\mathcal{O}(n\lg(n-k))$. The final step requires a formal derivative of polynomial degree $T$, and at most $2(n/T)$ times of $T$-point FFT. Thus, the complexity is $\mathcal{O}(n\lg(n-k))$. In summary, the proposed decoding algorithm requires $\mathcal{O}(n\lg(n-k)+(n-k)\lg^2(n-k))$.

\section{Concluding remarks}\label{sec:conclusion}
In the simulations, we implemented the algorithm in C and compiled it in 64-bit GCC compiler on Intel Xeon X5650 and Windows 7 platform. For $(n,k)=(2^{16}, 2^{15})$ RS codes over $\mathbb{F}_{2^{16}}$, the program took about $2.22\times 10^{-3}$ second to produce a codeword. We tested a codeword with $(n-k)/2$ errors, and the decoding takes about $0.401$ seconds. As for a comparison, we also ran the standard RS decoding algorithm~\cite{WinNT}, that took about $22.014$ seconds to decode a codeword. Thus, the proposed decoding is around $50$ times faster than the traditional approach under the parameter configurations described above. In our simulations, the proposed RS algorithm is suitable for long RS codes. 

In this paper, we developed fast decoding algorithms for $(n=2^m,k)$ systematic Reed-Solomon~(RS) codes over fields $\mathbb{F}_{2^m},m\in \mathbb{Z}^+$. The proposed algorithms are formed on a new basis $\mathbb{X}$~\cite{4319038}. We reformulated the formulas of the syndrome-based decoding algorithm, such that the FFTs for the new basis can be applied. Further, the fast polynomial division algorithm is proposed. We made some modifications such that the Newton iteration can be applied to the new basis. The fast Euclidean algorithm was also given in this paper. Combining these algorithms, a fast RS decoding algorithm is proposed, to achieve the complexity $\mathcal{O}(n\lg(n-k)+(n-k)\lg^2(n-k))$. By letting $k/n$ a constant, the complexity can be written as $\mathcal{O}(n\lg^2(n))$, that improves upon the best currently available decoding complexity of $\mathcal{O}(n\lg^2(n)\lg\lg(n))$~\cite{Gao02anew}. Although Justesen~\cite{1055516} had given the algorithm with the same complexity in 1976, it does not include the field $\mathbb{F}_{2^m}$, that can be recognized as the most important case in the real applications. 

The following we address some potential future works: 1.  To remove the constraint $(n-k)$ a power of two in the encoding/decoding algorithms. This will increase the  values of $n$ and $k$ to be selected. 2. To
generalize the algorithm to handle both errors and erasures. 3. To reduce the leading constant of the FFT approach. This will make the proposed algorithm more competitive for short codes.

\bibliographystyle{IEEEtran}
\bibliography{IEEEabrv,refs}
\newpage
\appendices

\section{Polynomial muliplication and formal derivative on new basis}\label{sec:operation2}
\cite{4319038} showed the polynomial multiplication and formal derivative in $\mathbb{\bar{X}}$. We take the similar procedure to show the corresponding operations in $\mathbb{X}$.
\subsection{Muliplication}
To multiply two polynomials, there exists a well-known fast approach based on FFT techniques. This approach can also be applied on the basis $\mathbb{X}$ over finite fields $\mathbb{F}_{2^m}$. Let $a(x)=\sum_{i=0}^{h-1} a_i\cdot X_i(x)$ and $b(x)=\sum_{i=0}^{h-1} b_i\cdot X_i(x)$ denote the two polynomials in $\mathbb{X}$. Its product $a(x)\cdot b(x)\pmod {s_i(x)}$ can be computed as
\[
\mathrm{IFFT}_\mathbb{X}(\mathrm{FFT}_\mathbb{X}(a, \beta)\otimes \mathrm{FFT}_\mathbb{X}(b, \beta), \beta),
\]
where $a=(a_0, a_1,\dots ,a_{h-1},0,\dots ,0)$ is a $2^i$-point vector represents the coefficients of $a(x)$ up to degree $2^i-1$. Similarly, $b$ is defined accordingly. The operation $\otimes$ performs pairwise multiplication on two vectors. This requires one $2^i$-point IFFT, two $2^i$-point FFTs and $2^i$ multiplications, and thus the complexity is $\mathcal{O}(2^i\lg(2^i))$.

\subsection{Formal derivative}
For a polynomial $D_{2^k}(x)$ in $\mathbb{X}$, we have
\begin{equation}\label{eq:derivative} 
\begin{aligned}
&D_{2^k}(x)=\sum_{i=0}^{2^k-1}d_iX_i(x)\\
=&\sum_{i=0}^{2^{k-1}-1}d_iX_i(x)+\sum_{i=2^{k-1}}^{2^k-1}d_iX_i(x)\\
=&\sum_{i=0}^{2^{k-1}-1}d_iX_i(x)+s_{k-1}(x)\sum_{i=0}^{2^{k-1}-1}d_{i+2^{k-1}}X_i(x)\\
=&D_{2^{k-1}}^{(0)}(x)+s_{k-1}(x)D_{2^{k-1}}^{(1)}(x).
\end{aligned}
\end{equation}
The formal derivative of $D_{2^k}(x)$ is given by
\begin{equation}\label{eq:derivative2}
\begin{aligned}
D_{2^k}'(x)=&[D_{2^{k-1}}^{(0)}]'(x)+s_{k-1}'(x)D_{2^{k-1}}^{(1)}(x)\\
&+s_{k-1}(x)[D_{2^{k-1}}^{(1)}]'(x).
\end{aligned}
\end{equation}
From Theorem \ref{canonical_form_W_i(x)}, $s'_{k-1}(x)$ is a constant. $[D_{2^{k-1}}^{(0)}]'(x)$ and $s_{k-1}(x)[D_{2^{k-1}}^{(1)}]'(x)$ can be computed recursively. Let $h=2^k$, and the recursive form of the complexity is written by $T(h)=2\cdot T(h/2)+\mathcal{O}(h)$ and then $T(h)=\mathcal{O}(h\lg (h))$.

\section{Proof of Lemmas}
\label{lemma-proof}
\subsection{Proof of Lemma \ref{MultipointEvaluation}}
\begin{proof}
From the definition, $\bar{D}_{2^k}(x)$ can be reformulated as
\begin{equation}\label{eqn_db0} 
\begin{aligned}
&\bar{D}_{2^k}(x)=\sum_{i=0}^{2^k-1}\bar{d}_i\bar{X}_i(x)\\
=&\sum_{i=0}^{2^{k-1}-1}\bar{d}_i\bar{X}_i(x)+\sum_{i=2^{k-1}}^{2^k-1}\bar{d}_i\bar{X}_i(x)\\
=&\sum_{i=0}^{2^{k-1}-1}\bar{d}_i\bar{X}_i(x)+\frac{s_{k-1}(x)}{s_{k-1}(v_{k-1})}\sum_{i=0}^{2^{k-1}-1}\bar{d}_{i+2^{k-1}}\bar{X}_i(x)\\
=&\sum_{i=0}^{2^{k-1}-1}(\bar{d}_i+\frac{s_{k-1}(x)}{s_{k-1}(v_{k-1})}\bar{d}_{i+2^{k-1}})\bar{X}_i(x).
\end{aligned}
\end{equation}
From Theorem \ref{canonical_form_W_i(x)}, given $\gamma \in \mathbb{F}_{2^m}$, we have
\begin{equation}\label{eq:n_db00} 
\begin{aligned}
s_{k-1}(a+\gamma)=&s_{k-1}(a)+s_{k-1}(\gamma)\\
=&s_{k-1}(\gamma)\qquad \forall a \in V_{k-1}.
\end{aligned}
\end{equation}
From \eqref{eqn_db0} and \eqref{eq:n_db00}, we have
\begin{equation}
\begin{aligned}
&\bar{D}_{2^k}(a+\gamma)\\
=&\sum_{i=0}^{2^{k-1}-1}(\bar{d}_i+\frac{s_{k-1}(a+\gamma)}{s_{k-1}(v_{k-1})}\bar{d}_{i+2^{k-1}})\bar{X}_i(a+\gamma)\\
=&\sum_{i=0}^{2^{k-1}-1}(\bar{d}_i+\frac{s_{k-1}(\gamma)}{s_{k-1}(v_{k-1})}\bar{d}_{i+2^{k-1}})\bar{X}_i(a+\gamma),
\end{aligned}
\end{equation}
for each $a\in V_{k-1}$. This completes the proof.
\end{proof}

\subsection{Proof of Lemma \ref{lemmaDa}}
\begin{proof}
From \eqref{eq:D_a}, we have
\begin{equation}\label{eq:a1}
\begin{aligned}
&D_a\\
=& \lceil\lg(\deg(a(x))+y+1)\rceil \\
=& \lceil\lg(\deg(a(x))+2^{D_\ell }-\deg(b(x)))\rceil \\
\geq &\lceil\lg(2^{D_\ell }+1)\rceil &\text{(From \eqref{eq:con})}\\
=& D_\ell +1.
\end{aligned}
\end{equation}
Moreover,
\begin{equation}\label{eq:a2}
\begin{aligned}
D_a=&\lceil\lg(\deg(a(x))+2^{D_\ell }-\deg(b(x)))\rceil \\
\leq &\lceil\lg(\deg(a(x))+2^{D_\ell })\rceil \\
\leq &\lceil\lg(2^{D_\ell }+2^{D_\ell })\rceil = D_\ell +1.
\end{aligned}
\end{equation}
\eqref{eq:a1}\eqref{eq:a2} concludes that $D_a=D_\ell +1$. This completes the proof.
\end{proof}

\subsection{Proof of Lemma \ref{lemmaQx}}
\begin{proof}
\eqref{eq:lemmaQx} is a summation of two terms. For the first term, we have
\begin{equation}
\begin{aligned}
&\deg(Q(x)\cdot H(x))\\
\leq &\deg(Q(x))+\deg(B(x))+1 &\text{(From \eqref{eq:LambdamathbbX(x)1})}\\
=&\deg(A(x))+1&\text{(From \eqref{eq:A_mathbbX(x)-1})}\\
=&\deg(a(x))+y+1&\text{(From \eqref{eq:A_mathbbX(x)3})}\\
=&\deg(a(x))+2^{D_\ell }-\deg(b(x))&\text{(From \eqref{eq:y})}\\
\leq &\deg(a(x))+2^{D_\ell }\\
\leq &2^{D_\ell }-1+2^{D_\ell }&\text{(From \eqref{eq:Dell})}\\
= &2^{D_a }-1.
\end{aligned}
\end{equation}
For the second term, we have
\begin{equation}
\begin{aligned}
&\deg(R(x)\cdot \Lambda(x)\cdot s_1(x))\\
=&\deg(r(x))+y+2^{D_\ell}+1\\
=&\deg(r(x))+(2^{D_\ell }-\deg(b(x))-1)+2^{D_\ell}+1&\text{(From \eqref{eq:y})}\\
=&2^{D_a }+\deg(r(x))-\deg(b(x))\\
\leq &2^{D_a}-1.
\end{aligned}
\end{equation}
This completes the proof.
\end{proof}

\subsection{Proof of Lemma \ref{lemma4}}
\begin{proof}
The proof follows mathematical induction. For the based case $i=0$, \eqref{eq:lambda_0(x)} shows the following holds.
\[
\Lambda_0(x)\cdot B_0(x)\cdot s_1(x)=s_1(x),
\]
and $\bar{r}_0(x)=0$.

Assume \eqref{eq:lambda(i)} holds at $i=j$. That is, 
\begin{equation}\label{eq:Lambdaj(x)}
\Lambda_j(x)\cdot B_j(x)\cdot s_1(x)=s_{j+1}(x)+\bar{r}_j(x),
\end{equation}
which is multiplied by $(s_j(x))^2$ to get
\begin{equation}\label{eq:assumation}
\begin{aligned}
&(s_j(x))^2\cdot \Lambda_j(x)\cdot B_j(x)\cdot s_1(x)\\
=&(s_j(x))^2\cdot s_{j+1}(x)+(s_j(x))^2\cdot \bar{r}_j(x).
\end{aligned}
\end{equation}
From \eqref{eq:w_j2(x)-1}, we have
\begin{equation}\label{eq:assumation1}
\begin{aligned}
&(s_j(x))^2\cdot s_{j+1}(x)+(s_j(x))^2\cdot \bar{r}_j(x)\\
=&(s_{j+1}(x))^2+s_j(v_j)s_j(x)s_{j+1}(x)+(s_j(x))^2\cdot \bar{r}_j(x)\\
=&s_{j+2}(x)+s_{j+1}(v_{j+1})s_{j+1}(x)+s_j(v_j)s_j(x)s_{j+1}(x)\\
&+(s_j(x))^2\cdot \bar{r}_j(x).
\end{aligned}
\end{equation}
By \eqref{eq:assumation1}\eqref{eq:B_i(x)3}, \eqref{eq:assumation} can be rewritten as
\begin{equation}\label{eq:assumation2}
\begin{aligned}
s_j(x)\cdot \Lambda_j(x)\cdot B_{j+1}(x)\cdot s_1(x)=s_{j+2}(x)+\hat{r}_j(x),
\end{aligned}
\end{equation}
where 
\begin{equation}
\begin{aligned}
\hat{r}_j(x)=&s_{j+1}(v_{j+1})s_{j+1}(x)+s_j(v_j)s_j(x)s_{j+1}(x)\\
&+(s_j(x))^2\cdot \bar{r}_j(x)\\
&+s_j(x)\cdot \Lambda_j(x)\cdot \bar{B}_j(x)\cdot s_1(x).
\end{aligned}
\end{equation}
The degree of each term of $\hat{r}_j(x)$ is
\[
\deg (s_{j+1}(v_{j+1})s_{j+1}(x))=2^{j+1},
\]
\[
\deg (s_j(v_j)s_j(x)s_{j+1}(x))=2^j+2^{j+1},
\]
\[
\deg ((s_j(x))^2\cdot \bar{r}_j(x))\leq 2^{j+1}+2^j,
\]
\begin{align*}
&\deg (s_j(x)\cdot \Lambda_j(x)\cdot \bar{B}_j(x)\cdot s_1(x))\\\leq &2^j+(2^j-1)+(2^j-1)+2.
\end{align*}
Thus, we have $\deg (\hat{r}_j(x))\leq 2^{j+1}+2^j$.

When $i=j+1$, from \eqref{eq:lambda(i)(x)}\eqref{eq:lambda(i)'(x)}, we have
\[
\Lambda_{j+1}(x)=\mathbb{Q}((s_j(x)\cdot \Lambda_j(x))^2\cdot B_{j+1}(x)\cdot s_1(x), j+2).
\]
The above equation can be rewritten as
\begin{equation}\label{eq:barLambda_j+1(x)}
\begin{aligned}
&\Lambda_{j+1}(x)\cdot s_{j+2}(x)+\check{r}_{j+2}(x)\\
=&(s_j(x)\cdot \Lambda_j(x))^2\cdot B_{j+1}(x)\cdot s_1(x),
\end{aligned}
\end{equation}
where $\deg (\check{r}_{j+2}(x))\leq 2^{j+2}-1$. We then  multiply \eqref{eq:barLambda_j+1(x)} by $B_{j+1}(x)\cdot s_1(x)$ to obtain
\begin{equation}\label{eq:barLambda_j+2(x)}
\begin{aligned}
&\Lambda_{j+1}(x)\cdot B_{j+1}(x)\cdot s_1(x)\cdot s_{j+2}(x)\\
&+\check{r}_{j+2}(x)\cdot B_{j+1}(x)\cdot s_1(x)\\
=&(s_j(x)\cdot\Lambda_j(x)\cdot B_{j+1}(x)\cdot s_1(x))^2\\
=&(s_{j+2}(x)+\hat{r}_j(x))^2 &\text{(By \eqref{eq:assumation2})}\\
=&(s_{j+2}(x))^2+(\hat{r}_j(x))^2.
\end{aligned}
\end{equation}
\eqref{eq:barLambda_j+2(x)} is then divided by $s_{j+2}(x)$ to get
\begin{equation}\label{eq:barLambda_j+3(x)}
\Lambda_{j+1}(x)\cdot B_{j+1}(x)\cdot s_1(x)=s_{j+2}(x)+\bar{r}_{j+1}(x),
\end{equation}
where
\[
\bar{r}_{j+1}(x)=\frac{(\hat{r}_j(x))^2-\check{r}_{j+2}(x)\cdot B_{j+1}(x)\cdot s_1(x)}{s_{j+2}(x)}.
\]
In \eqref{eq:barLambda_j+3(x)}, the degree of each term of $\deg (\bar{r}_{j+1}(x))$ is as follows:
\begin{align*}
&\deg ((\hat{r}_j)^2(x))\leq 2\cdot (2^{j+1}+2^j),\\
&\deg (\check{r}_{j+2}(x)\cdot B_{j+1}(x)\cdot s_1(x))\\
\leq& (2^{j+2}-1)+(2^{j+1}-1)+2,\\
&\deg (s_{j+2}(x))=2^{j+2}.
\end{align*}
Thus, $\deg (\bar{r}_{j+1}(x))\leq 2^{j+1}$. This completes the proof.
\end{proof}

\subsection{Proof of Lemma \ref{lemma5}}
\begin{proof}
From \eqref{eq:lambda(i)'(x)}, we have
\begin{equation}\label{eq:lambda(i)'(x)111}
\begin{aligned}
&\Lambda_i(x)\\
=&\mathbb{Q}((s_{i-1}(x))^2\cdot \bar{\Lambda}_i(x), i+1)\\
=&\mathbb{Q}((s_i(x)+s_{i-1}(v_{i-1})s_{i-1}(x))
\cdot \bar{\Lambda}_i(x), i+1)&\text{(From \eqref{eq:w_j2(x)-1})}\\
=&\mathbb{Q}(s_i(x)\bar{\Lambda}_i(x), i+1)\\
&+s_{i-1}(v_{i-1})\cdot \mathbb{Q}(s_{i-1}(x)\bar{\Lambda}_i(x), i+1)\\
\end{aligned}
\end{equation}
\eqref{eq:lambda(i)'(x)111} has two terms, and we recalled that $\deg(\bar{\Lambda}_i(x))=2^{i+1}-1$. Let
\[
\bar{\Lambda}_i(x)=\Lambda_{i}^{(0)}(x)+s_i(x)\Lambda_{i}^{(1)}(x),
\]
where both $\Lambda_{i}^{(0)}(x)$ and $\Lambda_{i}^{(1)}(x)$ have degrees no more than $2^i-1$. Then
\begin{equation}\label{eq:barlambda(i)(x)}
\begin{aligned}
&s_i(x)\bar{\Lambda}_i(x)\\
=&s_i(x)(\Lambda_{i}^{(0)}(x)+s_i(x)\Lambda_{i}^{(1)}(x))\\
=&s_i(x)\Lambda_{i}^{(0)}(x)+(s_i(x))^2\Lambda_{i}^{(1)}(x)\\
=&s_i(x)\Lambda_{i}^{(0)}(x)+s_i(v_i)s_i(x)\Lambda_{i}^{(1)}(x)\\
&+s_{i+1}(x)\Lambda_{i}^{(1)}(x).
\end{aligned}
\end{equation}
From \eqref{eq:barlambda(i)(x)}, the first term in \eqref{eq:lambda(i)'(x)111} can be reformulated as
\[
\mathbb{Q}(s_i(x)\bar{\Lambda}_i(x), i+1)=\Lambda_{i}^{(1)}(x)=\mathbb{Q}(\bar{\Lambda}_i(x), i).
\]
With the similar step, it can be shown that the second stem can be formulated as
\begin{align*}
&s_{i-1}(v_{i-1})\mathbb{Q}(s_{i-1}(x)\bar{\Lambda}_i(x), i+1)\\
=&\mathbb{Q}(\Lambda_i^{(1)}(x), i-1)\cdot s_{i-1}(v_{i-1}).
\end{align*}
This completes the proof.
\end{proof}

\subsection{Proof of Lemma~\ref{lemma6.1}}
\begin{proof}

 For the based case $\deg(b(x))<2^{g-1}$~(see Algorithm~\ref{alg:HalfGCD}, line 1), it is clear that~\eqref{eq:Eucint} holds.

Assume Algorithm~\ref{alg:HalfGCD} is valid for $\mathrm{HGCD}(a(x), b(x), j)$ with $j\leq g-1$. When $j=g$, the degree of $a(x)$ is between $2^{g-1}\leq \deg(a(x))\leq 2^g-1$. In this case, both $a(x)$ and $b(x)$ are divided into three individual polynomials as expressed in \eqref{eq:axdivid2} and \eqref{eq:axdivide3}. In line 3, $\mathrm{HGCD}(a_\mathrm{H}(x), b_\mathrm{H}(x), g-1)$ is called to obtain $(Z_\mathrm{H}, M_\mathrm{H})$, that possesses
\begin{equation}\label{eq:lem1}
Z_\mathrm{H}=M_\mathrm{H}\cdot \begin{bmatrix}
a_\mathrm{H}(x)\\ 
b_\mathrm{H}(x)
\end{bmatrix}.
\end{equation}
Multiplying \eqref{eq:lem1}  by $s_{g-1}(x)$ to obtain
\begin{equation}\label{eq:lem2}
Z_\mathrm{H}\cdot s_{g-1}(x)=M_\mathrm{H}\cdot s_{g-1}(x)\begin{bmatrix}
a_\mathrm{H}(x)\\ 
b_\mathrm{H}(x)
\end{bmatrix}\\
\end{equation}
which is equivalent to
\begin{equation}\label{eq:lem3}
\begin{aligned}
&Z_\mathrm{H}\cdot s_{g-1}(x)+
M_\mathrm{H}
\begin{bmatrix}
a_\mathrm{L}(x)\\ 
b_\mathrm{L}(x)
\end{bmatrix}\\
=& M_\mathrm{H}
\begin{bmatrix}
s_{g-1}(x)\cdot a_\mathrm{H}(x)+a_\mathrm{L}(x)\\ 
s_{g-1}(x)\cdot b_\mathrm{H}(x)+b_\mathrm{L}(x)
\end{bmatrix}.
\end{aligned}
\end{equation}
By \eqref{eq:axdivid2} and \eqref{eq:axdivide3}, \eqref{eq:lem3} becomes
\begin{equation}
Z_\mathrm{H}\cdot s_{g-1}(x)+
M_\mathrm{H}
\begin{bmatrix}
a_\mathrm{L}(x)\\ 
b_\mathrm{L}(x)
\end{bmatrix}
= M_\mathrm{H}
\begin{bmatrix}
a(x)\\ 
b(x)
\end{bmatrix}.
\end{equation}
Then we have
\begin{equation}\label{eq:lem4}
\begin{bmatrix}
z_\mathrm{M0}(x)\\ 
z_\mathrm{M1}(x)
\end{bmatrix}
=M_\mathrm{H}
\begin{bmatrix}
a(x)\\ 
b(x)
\end{bmatrix}
\mbox{ and }
Z_\mathrm{M}
=M_\mathrm{H}
\begin{bmatrix}
a(x)\\ 
b(x)
\end{bmatrix}.
\end{equation}
Note that $z_\mathrm{M0}(x)$ and $z_\mathrm{M1}(x)$ are computed in line 4.  \eqref{eq:lem4} shows that $(Z_\mathrm{M}, M_\mathrm{H})$ satisfies the equality, and thus the return in line 5 is valid.

In line 7, $z_\mathrm{M0}(x)$ is divided by $z_\mathrm{M1}(x)$ to get
\begin{equation}\label{eq:dividEuclid11}
z_\mathrm{M0}(x)=q_\mathrm{M}(x)\cdot z_\mathrm{M1}(x)+r_\mathrm{M}(x),
\end{equation}
with 
\begin{equation}\label{eq:dividEuclid12}
\deg(r_\mathrm{M}(x))<\deg(z_\mathrm{M1}(x)).
\end{equation}
The matrix form of \eqref{eq:dividEuclid11} can be reformulated as
\begin{equation}\label{eq:dividEuclid11m}
\begin{bmatrix}
z_\mathrm{M1}(x)\\ 
r_\mathrm{M}(x)
\end{bmatrix}
=
\begin{bmatrix}
0 & 1\\ 
1 & -q_\mathrm{M}(x)
\end{bmatrix}
\begin{bmatrix}
z_\mathrm{M0}(x)\\ 
z_\mathrm{M1}(x)
\end{bmatrix}.
\end{equation}
Then $z_\mathrm{M1}(x)$ and $r_\mathrm{M}(x)$ are decomposed into several  polynomials as
\begin{equation}\label{eq:axdivide0}
\begin{aligned}
&z_\mathrm{M1}(x)\\
=&z_\mathrm{M1LL}(x)+s_{g-2}(x)z_\mathrm{M1LH}(x)+s_{g-1}(x)z_\mathrm{M1H}(x)\\
=&z_\mathrm{M1LL}(x)+s_{g-2}(x)z_\mathrm{M1LH}(x)\\
&+s_{g-2}(x)(s_{g-2}(x)+s_{g-2}(v_{g-2}))z_\mathrm{M1H}(x)\\
=&z_\mathrm{M1LL}(x)+s_{g-2}(x)z_\mathrm{M1M}(x),
\end{aligned}
\end{equation}
where
\begin{equation}\label{eq:axdivide00}
\begin{aligned}
&z_\mathrm{M1M}(x)\\
=&z_\mathrm{M1LH}(x)+(s_{g-2}(x)+s_{g-2}(v_{g-2}))z_\mathrm{M1H}(x).
\end{aligned}
\end{equation}
Similarly, 
\begin{equation}\label{eq:axdivide1}
r_\mathrm{M}(x)
=r_\mathrm{MLL}(x)+s_{g-2}(x)r_\mathrm{MM}(x),
\end{equation}
where 
\begin{equation}\label{eq:axdivide01}
\begin{aligned}
&r_\mathrm{MM}(x)\\
=&r_\mathrm{MLH}(x)+(s_{g-2}(x)+s_{g-2}(v_{g-2}))r_\mathrm{MH}(x).
\end{aligned}
\end{equation}
$z_\mathrm{M1M}(x)$ (and $r_\mathrm{MM}(x)$) can be treated as the quotient of dividing $z_\mathrm{M1}(x)$ (and $r_\mathrm{M}(x)$) by $s_{g-2}(x)$. By \eqref{eq:dividEuclid12}, this implies
\[
\deg(r_\mathrm{MM}(x))<\deg(z_\mathrm{M1M}(x)).
\]

Line 9 calls $\mathrm{HGCD}(z_\mathrm{M1M}(x), r_\mathrm{MM}(x), g-1)$ to obtain $(Y_\mathrm{M},M_\mathrm{M})$ possessing
\begin{equation}\label{eq:lem5}
\begin{aligned}
Y_\mathrm{M}
=M_\mathrm{M}
\begin{bmatrix}
z_\mathrm{M1M}(x)\\ 
r_\mathrm{MM}(x)
\end{bmatrix}.
\end{aligned}
\end{equation}
Multiplying \eqref{eq:lem5}  by $s_{g-2}(x)$ to obtain
\begin{equation}\label{eq:lem59}
Y_\mathrm{M}\cdot s_{g-2}(x)
=M_\mathrm{M}\cdot s_{g-2}(x)
\begin{bmatrix}
z_\mathrm{M1M}(x)\\ 
r_\mathrm{MM}(x).
\end{bmatrix}
\end{equation}
By adding $\begin{bmatrix}
z_\mathrm{M1LL}(x)\\ 
r_\mathrm{MLL}(x)
\end{bmatrix}
$ to both side of \eqref{eq:lem59}, we have
\begin{equation}
\begin{aligned}
&Y_\mathrm{M}\cdot s_{g-2}(x)+
M_\mathrm{M}
\begin{bmatrix}
z_\mathrm{M1LL}(x)\\ 
r_\mathrm{MLL}(x)
\end{bmatrix}\\
=&M_\mathrm{M}
\begin{bmatrix}
z_\mathrm{M1LL}(x)+z_\mathrm{M1M}(x)\cdot s_{g-2}(x)\\ 
r_\mathrm{MLL}(x)+r_\mathrm{MM}(x)\cdot s_{g-2}(x)
\end{bmatrix}
\end{aligned}
\end{equation}
which is equivalent to 
\begin{equation}\label{eq:lem55}
Y_\mathrm{M}\cdot s_{g-2}(x)+
M_\mathrm{M}
\begin{bmatrix}
z_\mathrm{M1LL}(x)\\ 
r_\mathrm{MLL}(x)
\end{bmatrix}
=M_\mathrm{M}
\begin{bmatrix}
z_\mathrm{M1}(x)\\ 
r_\mathrm{M}(x)
\end{bmatrix}.
\end{equation}
Substituting \eqref{eq:dividEuclid11m} and \eqref{eq:lem4} into \eqref{eq:lem55} to obtain
\begin{equation}\label{eq:lem6}
\begin{aligned}
&Y_\mathrm{M}\cdot s_{g-2}(x)+
M_\mathrm{M}
\begin{bmatrix}
z_\mathrm{M1LL}(x)\\ 
r_\mathrm{MLL}(x)
\end{bmatrix}\\
=&M_\mathrm{M}
\begin{bmatrix}
0 & 1\\ 
1 & -q_\mathrm{M}(x)
\end{bmatrix}
M_\mathrm{H}
\begin{bmatrix}
a(x)\\ 
b(x)
\end{bmatrix}.
\end{aligned}
\end{equation}
Hence, we have
\begin{equation}
 Z_\mathrm{R}
=M_\mathrm{R}
\begin{bmatrix}
a(x)\\ 
b(x)
\end{bmatrix},
\end{equation}
where
\begin{equation}\label{eq:Mr}
M_\mathrm{R}=M_\mathrm{M}
\begin{bmatrix}
0 & 1\\ 
1 & -q_\mathrm{M}(x)
\end{bmatrix}
M_\mathrm{H}
\end{equation}
and
\begin{equation}\label{eq:Zr}
Z_\mathrm{R}=
Y_\mathrm{M}\cdot s_{g-2}(x)+
M_\mathrm{M}
\begin{bmatrix}
z_\mathrm{M1LL}(x)\\ 
r_\mathrm{MLL}(x)
\end{bmatrix}
\end{equation}
are the return results in Line 10. 
\end{proof}

\subsection{Proof of Lemma~\ref{lemma6.2}}
\begin{proof}
Assume $\mathrm{HGCD}(a(x), b(x), i)$ is valid for $i\leq g-1$, i.e.,  the recursive calls in line 3~(and line 9) are valid. Assume $i=q$. It is clear that the call at line 3 satisfies the condition, since $a_\mathrm{H}(x)$ and $b_\mathrm{H}(x)$ are the high degree portions of $a(x)$ and $b(x)$, respectively. 

For the call at line 9, we first consider the degree of $z_\mathrm{M1}(x)$. For simplicity, $(Z_\mathrm{H},M_\mathrm{H})$ is denoted as
\[
Z_\mathrm{H}=
\begin{bmatrix}
z_\mathrm{H0}(x)\\ 
z_\mathrm{H1}(x)
\end{bmatrix},
M_\mathrm{H}=
\begin{bmatrix}
m_\mathrm{H00}(x)&m_\mathrm{H01}(x)\\ 
m_\mathrm{H10}(x)&m_\mathrm{H11}(x)
\end{bmatrix}.
\]
Because $\deg(a_\mathrm{L}(x))\leq 2^{q-2}-1$, $\deg(b_\mathrm{L}(x))\leq 2^{q-2}-1$, and $\deg(z_\mathrm{H1}(x))\leq 2^{q-2}-1$, from the assumption, we have
\begin{equation}\label{eq:degzm1}
\begin{aligned}
\deg (z_\mathrm{M1}(x))\leq \max\{&(2^{q-2}-1)+2^{g-1}, \\
&\deg(m_\mathrm{H10}(x))+(2^{q-2}-1), \\
&\deg(m_\mathrm{H11}(x))+(2^{q-2}-1)\}.
\end{aligned}
\end{equation}
From the assumption, 
\[
\deg(m_\mathrm{H10}(x))\leq\deg(m_\mathrm{H11}(x))\leq \deg(a_\mathrm{H}(x))-\deg(z_\mathrm{H0}(x)).
\]
As $\deg(a_\mathrm{H}(x))\leq 2^{q-1}-1$ and $\deg(z_\mathrm{H0}(x))\geq 2^{q-2}$, we have
\[
\deg(m_\mathrm{H10}(x))\leq\deg(m_\mathrm{H11}(x))\leq 2^{q-2}-1.
\]
Then \eqref{eq:degzm1} gives
\begin{equation}\label{eq:degzm12}
\begin{aligned}
&\deg (z_\mathrm{M1}(x))\\
\leq &\max\{(2^{g-2}-1)+2^{g-1}, (2^{q-2}-1)+(2^{q-2}-1)\}\\
=&(2^{g-2}-1)+2^{g-1}.
\end{aligned}
\end{equation}
Thus, the inequality
\[
\deg(r_\mathrm{M}(x))<\deg(z_\mathrm{M1}(x))\leq 2^{g-1}+2^{g-2}-1
\]
in line 7 is valid. In line 8, $z_\mathrm{M1M}(x)$ and $r_\mathrm{MM}(x)$ are the quotients of dividing $r_\mathrm{M}(x)$ and $z_\mathrm{M1}(x)$ by $s_{g-2}(x)$, and then
\begin{equation}\label{eq:degzm222}
\deg(r_\mathrm{MM}(x))<\deg(z_\mathrm{M1M}(x))\leq 2^{g-1}-1.
\end{equation}
Further, due to the if condition in line 5, we have
\[
\deg(z_\mathrm{M1}(x))\geq 2^{g-1}
\]
after line 7. This implies
\begin{equation}\label{eq:degzm220}
\deg(z_\mathrm{M1M}(x))\geq 2^{g-2}.
\end{equation}
By \eqref{eq:degzm222} and \eqref{eq:degzm220}, the requirements of the call in line 9 are verified.
\end{proof}

\subsection{Proof of Lemma \ref{lemma6.5}}
\begin{proof}
Algorithm~\ref{alg:HalfGCD} has three returns at lines 1, 5 and 10. Assume that the recursive call HGCD in line 3 and line 10 outputs the valid results.  For line 1, it is clear to see it. For line 5, \eqref{eq:lem2}-\eqref{eq:lem4} show that the degree of $z_\mathrm{M0}(x)$ is at least
\[
\deg(z_\mathrm{M0}(x))\geq \deg(z_\mathrm{H0})+2^{g-1}.
\]
By the assumption,  $\deg(z_\mathrm{H0})\geq 2^{g-2}$ and we have
\begin{equation}\label{eq:zm10}
\deg(z_\mathrm{M0}(x))\geq 2^{g-2}+2^{g-1}.
\end{equation}
By \eqref{eq:zm10} and the if condition in line 5, the first condition holds.

Let us consider line 10.  $Z_\mathrm{R}$, $Y_\mathrm{M}$ and $M_\mathrm{M}$ can be denoted as
\begin{align*}
Z_\mathrm{R}=&
\begin{bmatrix}
z_\mathrm{R0}(x)\\ 
z_\mathrm{R1}(x)
\end{bmatrix},
Y_\mathrm{M}=
\begin{bmatrix}
y_\mathrm{M0}(x)\\ 
y_\mathrm{M1}(x)
\end{bmatrix},\\
M_\mathrm{M}=&
\begin{bmatrix}
m_\mathrm{M00}(x)&m_\mathrm{M01}(x)\\ 
m_\mathrm{M10}(x)&m_\mathrm{M11}(x)
\end{bmatrix}.
\end{align*}
The degree of $z_\mathrm{R0}(x)$ is at least
\begin{align*}
&\deg(z_\mathrm{R0}(x))\geq \deg(y_\mathrm{M0}(x))+2^{g-2}\\
\geq &2^{g-2}+2^{g-2}=2^{g-1}.
\end{align*}
Further,
\begin{equation}\label{eq:lem83}
\begin{aligned}
\deg(z_\mathrm{R1}(x))= \max\{&\deg(y_\mathrm{M1}(x))+2^{g-2}, \\
&\deg(m_\mathrm{M10}(x)z_\mathrm{M1LL}(x)),\\
&\deg(m_\mathrm{M11}(x)r_\mathrm{MLL}(x))\}.
\end{aligned}
\end{equation}
By assumptions, we have
\begin{align*}
&\deg(m_\mathrm{M10}(x))\leq \deg(m_\mathrm{M11}(x))\\
&\leq\deg(z_\mathrm{M1M}(x))-\deg(y_\mathrm{M0}(x))\\
&\leq (2^{q-1}-1)-2^{q-2}=2^{q-2}-1.
\end{align*}
Then \eqref{eq:lem83} becomes
\begin{equation}\label{eq:degzm2}
\begin{aligned}
&\deg (z_\mathrm{R1}(x))\\
\leq &\max\{(2^{g-2}-1)+2^{g-2}, (2^{q-2}-1)+(2^{q-2}-1)\}\\
=&2^{g-1}-1,
\end{aligned}
\end{equation}
as, by assumptions, $\deg(z_\mathrm{M1LL}(x))\leq 2^{q-2}-1$ and $\deg(r_\mathrm{MLL}(x))\leq 2^{q-2}-1$, and $\deg(y_\mathrm{M1}(x))\leq 2^{g-2}-1$. This verifies \eqref{eq:cond1}.
\end{proof}

\subsection{Proof of Lemma~\ref{lemma6.55}}
\begin{proof}
Assume that the recursive call HGCD in line 3 and line 10 outputs the valid results.  For the base case in line 1, it is clear that the condition holds. Notice that $\deg(0)$ is a special case, we can treat $\deg(0)=0$ in this case. For line 5, the objective is to prove
\begin{equation}\label{eq:line5c340}
\deg(m_\mathrm{H11}(x))\leq\deg(a(x))-\deg(z_\mathrm{M0}(x)),
\end{equation}
and
\begin{equation}\label{eq:line5c341}
\begin{aligned}
\deg(m_\mathrm{Hi0}(x))&\leq\deg(m_\mathrm{Hi1}(x)),\\
\deg(m_\mathrm{H0i}(x))&\leq\deg(m_\mathrm{H1i}(x)),\qquad i=0,1.\\
\end{aligned}
\end{equation}
By assumptions, line 3 of the algorithm gives
\begin{equation}\label{eq:mH11}
\deg(m_\mathrm{H11}(x))\leq\deg(a_\mathrm{H}(x))-\deg(z_\mathrm{H0}(x)),
\end{equation}
and
\begin{equation}\label{eq:mH110}
\begin{aligned}
\deg(m_\mathrm{Hi0}(x))&\leq\deg(m_\mathrm{Hi1}(x)),\\
\deg(m_\mathrm{H0i}(x))&\leq\deg(m_\mathrm{H1i}(x)),\qquad i=0,1.\\
\end{aligned}
\end{equation}
\eqref{eq:mH110} verifies that \eqref{eq:line5c341} is true. Further, \eqref{eq:lem2}-\eqref{eq:lem4} show that \eqref{eq:mH11} can be reformed as
\begin{equation}\label{eq:mH100}
\begin{aligned}
&\deg(m_\mathrm{H11}(x))\\
\leq&\deg(a_\mathrm{H}(x)s_{g-1}(x))-\deg(z_\mathrm{H0}(x)s_{g-1}(x))\\
= &\deg(a(x))-\deg(z_\mathrm{M0}(x)).
\end{aligned}
\end{equation}
This verifies \eqref{eq:line5c340}.

Let us consider line 10. $Z_\mathrm{R}$, $Y_\mathrm{M}$ $M_\mathrm{M}$, and $M_\mathrm{R}$ can be denoted as
\begin{align*}
&Z_\mathrm{R}=
\begin{bmatrix}
z_\mathrm{R0}(x)\\ 
z_\mathrm{R1}(x)
\end{bmatrix},
Y_\mathrm{M}=
\begin{bmatrix}
y_\mathrm{M0}(x)\\ 
y_\mathrm{M1}(x)
\end{bmatrix},\\
&M_\mathrm{M}=
\begin{bmatrix}
m_\mathrm{M00}(x)&m_\mathrm{M01}(x)\\ 
m_\mathrm{M10}(x)&m_\mathrm{M11}(x)
\end{bmatrix}.
M_\mathrm{R}=
\begin{bmatrix}
m_\mathrm{R00}(x)&m_\mathrm{R01}(x)\\ 
m_\mathrm{R10}(x)&m_\mathrm{R11}(x)
\end{bmatrix}.
\end{align*}
The objective is to prove
\begin{equation}\label{eq:line10c340}
\deg(m_\mathrm{R11}(x))\leq\deg(a(x))-\deg(z_\mathrm{R0}(x)),
\end{equation}
and
\begin{equation}\label{eq:line10c341}
\begin{aligned}
\deg(m_\mathrm{Ri0}(x))&\leq\deg(m_\mathrm{Ri1}(x)),\\
\deg(m_\mathrm{R0i}(x))&\leq\deg(m_\mathrm{R1i}(x)),\qquad i=0,1.\\
\end{aligned}
\end{equation}
By assumptions, line 9 of the algorithm gives
\begin{equation}\label{eq:mm11}
\deg(m_\mathrm{M11}(x))\leq\deg(z_\mathrm{M1M}(x))-\deg(y_\mathrm{M0}(x)),
\end{equation}
and
\begin{equation}\label{eq:mm111}
\begin{aligned}
\deg(m_\mathrm{Mi0}(x))&\leq\deg(m_\mathrm{Mi1}(x)),\\
\deg(m_\mathrm{M0i}(x))&\leq\deg(m_\mathrm{M1i}(x)),\qquad i=0,1.\\
\end{aligned}
\end{equation}
To verify \eqref{eq:line10c341}, \eqref{eq:Mr} can be reformed as
\begin{equation}\label{Mr2}
\begin{aligned}
&M_\mathrm{R}\\
=&
\begin{bmatrix}
m_\mathrm{M00}(x) & m_\mathrm{M01}(x)\\ 
m_\mathrm{M10}(x) & m_\mathrm{M11}(x)
\end{bmatrix}
\begin{bmatrix}
m_\mathrm{H10}(x) & m_\mathrm{H11}(x)\\ 
m_\mathrm{H00}(x) & m_\mathrm{H01}(x)
\end{bmatrix}\\
&-q_\mathrm{M}(x)
\begin{bmatrix}
m_\mathrm{M01}(x)\\
m_\mathrm{M11}(x)
\end{bmatrix}
\begin{bmatrix}
m_\mathrm{H10}(x)&
m_\mathrm{H11}(x)
\end{bmatrix}.
\end{aligned}
\end{equation}
Based on assumptions~\eqref{eq:mH110} and \eqref{eq:line10c341}, it can be seen that the degrees of the elements of $M_\mathrm{R}$ are determined by the second term. Precisely,
\begin{equation}\label{eq:Mr2d}
\begin{aligned}
\deg(m_\mathrm{R00}(x))&=\deg(q_\mathrm{M}(x))+\deg(m_\mathrm{M01}(x))+\deg(m_\mathrm{H10}(x)),\\
\deg(m_\mathrm{R01}(x))&=\deg(q_\mathrm{M}(x))+\deg(m_\mathrm{M01}(x))+\deg(m_\mathrm{H11}(x)),\\
\deg(m_\mathrm{R10}(x))&=\deg(q_\mathrm{M}(x))+\deg(m_\mathrm{M11}(x))+\deg(m_\mathrm{H10}(x)),\\
\deg(m_\mathrm{R11}(x))&=\deg(q_\mathrm{M}(x))+\deg(m_\mathrm{M11}(x))+\deg(m_\mathrm{H11}(x)).\\
\end{aligned}
\end{equation}
From the assumptions $\deg(m_\mathrm{M01}(x))\leq \deg(m_\mathrm{M11}(x))$ and $\deg(m_\mathrm{H10}(x))\leq \deg(m_\mathrm{H11}(x))$, \eqref{eq:line10c341} can be verified.

The verification of \eqref{eq:line10c340} is considered as follows. From \eqref{eq:axdivide01}, \eqref{eq:mm11} can be reformed as
\begin{equation}\label{eq:mm110}
\begin{aligned}
&\deg(m_\mathrm{M11}(x))\\
\leq&\deg(z_\mathrm{M1M}(x)s_{g-2}(x))-\deg(y_\mathrm{M0}(x)s_{g-2}(x))\\
=&\deg(z_\mathrm{M1}(x))-\deg(z_\mathrm{R0}(x)).\\
\end{aligned}
\end{equation}
\eqref{eq:mm110} is summed by \eqref{eq:mH100}, resulting in \eqref{eq:mm111v}, and thus \eqref{eq:line10c340} is verified.
\begin{figure*}[b]
\hrulefill
\begin{equation}\label{eq:mm111v}
\begin{aligned}
&\deg(m_\mathrm{M11}(x))+\deg(m_\mathrm{H11}(x))\leq \deg(z_\mathrm{M1}(x))-\deg(z_\mathrm{R0}(x))+\deg(a(x))-\deg(z_\mathrm{M0}(x))\\
\Rightarrow & \deg(m_\mathrm{M11}(x))+\deg(m_\mathrm{H11}(x))+\deg(z_\mathrm{M0}(x))-\deg(z_\mathrm{M1}(x))\leq \deg(a(x))-\deg(z_\mathrm{R0}(x))\\
\Rightarrow & \deg(m_\mathrm{M11}(x))+\deg(m_\mathrm{H11}(x))+\deg(q_\mathrm{M}(x))\leq \deg(a(x))-\deg(z_\mathrm{R0}(x))&\text(By~\eqref{eq:dividEuclid11})\\
\Rightarrow & \deg(m_\mathrm{R11}(x))\leq \deg(a(x))-\deg(z_\mathrm{R0}(x))&\text(By~\eqref{eq:Mr2d}),
\end{aligned}
\end{equation}
\end{figure*}
\end{proof}

\subsection{Proof of Lemma \ref{lemma7}}
\begin{proof}
The proof follows mathematical induction. We pick $T=n/2$ as the base case.
Then from \eqref{eq:u}, $\mathbf{u}=(\mathbf{u}_0,\mathbf{\omega_0})$ has $n/2$ $\omega_0$s in the high degree part. From \eqref{eq:mathbfv}, $\mathbf{v}=(\mathbf{v}_0,\mathbf{v}_1)$ is divided into two equal sub-vectors. Then \eqref{eq:encoding1} can be written as
\[
(\mathbf{u}_0,\mathbf{\omega_0})=\mathrm{IFFT}_\mathbb{\bar{X}}((\mathbf{v}_0,\mathbf{v}_1),m,\omega_0).
\]
In Algorithm~\ref{alg:IFFT}, Line 3 computes $D^{(0)}=\mathrm{IFFT}_\mathbb{\bar{X}}(\mathbf{v}_0, m-1,\omega_0)$, and Line 4 computes $D^{(1)}=\mathrm{IFFT}_\mathbb{\bar{X}}(\mathbf{v}_1, m-1,\omega_{n/2})$. The vector $\mathbf{\omega_0}$ is calculated by line 6, and $\mathbf{u}_0$ is computed by line 7. As line 6 only requires pointwise additions, which can be written as a vector addition:
\begin{equation}\label{eq:encoding20}
\begin{aligned}
&\mathbf{\omega_0}=D^{(0)}+D^{(1)}\\
=&\mathrm{IFFT}_\mathbb{\bar{X}}(\mathbf{v}_0, m-1,\omega_0)+\mathrm{IFFT}_\mathbb{\bar{X}}(\mathbf{v}_1, m-1,\omega_{n/2}).
\end{aligned}
\end{equation}

Assume \eqref{eq:encoding2} holds at $T=S=2^s$, and thus
\begin{equation}\label{eq:encoding22}
\begin{aligned}
\mathbf{\omega_0}=&\mathrm{IFFT}_\mathbb{\bar{X}}(\mathbf{v}_0, s,\omega_0)+\mathrm{IFFT}_\mathbb{\bar{X}}(\mathbf{v}_1, s,\omega_S)+\dots \\
&+\mathrm{IFFT}_\mathbb{\bar{X}}(\mathbf{v}_{n/S-1}, s, \omega_{n-S}).
\end{aligned}
\end{equation}
When $T=S/2=2^{s-1}$, \eqref{eq:encoding22}  becomes
\begin{equation}\label{eq:encoding23}
\begin{aligned}
&(\mathbf{u}_{n/(S/2)-2},\mathbf{\omega_0})\\
=&\mathrm{IFFT}_\mathbb{\bar{X}}(\mathbf{v}_0, s,\omega_0)+\mathrm{IFFT}_\mathbb{\bar{X}}(\mathbf{v}_1, s,\omega_S)+\dots \\
&+\mathrm{IFFT}_\mathbb{\bar{X}}(\mathbf{v}_{n/S-1}. s, \omega_{n-S}).
\end{aligned}
\end{equation}
We can extract the computations regarding $\mathbf{\omega_0}$ in \eqref{eq:encoding23}. Similarly, this decomposes each $s$-point IFFT into two $(s/2)$-point IFFTs, resulting in
\begin{equation}\label{eq:encoding24}
\begin{aligned}
\mathbf{\omega_0}=&\mathrm{IFFT}_\mathbb{\bar{X}}(\mathbf{v}_0, s-1,\omega_0)+\mathrm{IFFT}_\mathbb{\bar{X}}(\mathbf{v}_1, s-1,\omega_{S/2})+\dots \\
&+\mathrm{IFFT}_\mathbb{\bar{X}}(\mathbf{v}_{n/S-1}, s-1, \omega_{n-S/2}),
\end{aligned}
\end{equation}
This completes the proof.
\end{proof}

\end{document}